\documentclass[aps,prb,twocolumn,superscriptaddress,floatfix,10pt]{revtex4-2}
\usepackage{verbatim}
\usepackage{physics}
\usepackage{bbold}
\usepackage[percent]{overpic}
\usepackage{xcolor}
\usepackage{orcidlink}
\usepackage{graphicx}
\usepackage{amsmath,amssymb,amsfonts,nccmath}
\usepackage{bm}
\usepackage{mathtools}

\usepackage[caption=false]{subfig}
\usepackage{hyperref}
\hypersetup{
	colorlinks=true,
	linkcolor=blue,
	filecolor=blue,
	citecolor = blue,      
	urlcolor=blue,
}

\newcommand{\floor}[1]{\lfloor #1 \rfloor}
\DeclareMathOperator{\sgn}{sgn}
\newcommand{\krodelta}[1]{\delta_{#1}}
\newcommand{\identity}{\mathbb{1}}
\newcommand{\Msin}[1]{\sin(#1)}
\newcommand{\Mcos}[1]{\cos(#1)}
\newcommand{\Mcosh}[1]{\cosh(#1)}
\newcommand{\Msinh}[1]{\sinh(#1)}
\newcommand{\Mtan}[1]{\tan(#1)}
\newcommand{\Mtanh}[1]{\tanh(#1)}

\newcommand{\Msinq}[1]{\sin^2\left(#1\right)}

\newcommand{\sigmaz}{\hat{\sigma}^{z}}
\newcommand{\sigmax}{\hat{\sigma}^{x}}
\newcommand{\sigmay}{\hat{\sigma}^{y}}
\newcommand{\hvf}{\hbar v_{\rm F}}
\newcommand{\kBT}{k_{\rm B}T}
\newcommand{\kBTc}{k_{\rm B}T_{\rm c}}
\newcommand{\deltaO}{\Delta_{0}}

\newcommand{\kf}{k_{\rm F}}
\newcommand{\vphi}{\varphi}
\newcommand{\vphip}{\varphi^{\prime}}

\newcommand{\df}{\partial_{\vphi}}
\newcommand{\dqf}{\partial^2_{\vphi}}

\newcommand{\A}{\hat{a}}
\newcommand{\Ac}{\hat{a}^{\dag}}
\newcommand{\B}{\hat{b}}
\newcommand{\Bc}{\hat{b}^{\dag}}

\newcommand{\cdz}[1]{d_{z}(#1,\varphi)}
\newcommand{\cdx}[1]{d_{x}(#1,\varphi)}
\newcommand{\vd}[1]{\vb{d}(#1,\varphi)}

\newcommand{\gammadagup}[1]{\hat{\gamma}^{\dag}_{+,#1}}
\newcommand{\gammadagdw}[1]{\hat{\gamma}^{\dag}_{-,#1}}

\newcommand{\gammaopdag}[1]{\hat{\gamma}^{\dag}_{#1}}
\newcommand{\gammaop}[1]{\hat{\gamma}^{\mathstrut}_{#1}}

\newcommand{\PSidag}[1]{\hat{\psi}^{\dag}_{#1}}
\newcommand{\PSi}[1]{\hat{\psi}^{\mathstrut}_{#1}}

\newcommand{\Psidagup}[1]{\hat{\psi}^{\dag}_{+,#1}}
\newcommand{\Psidagdw}[1]{\hat{\psi}^{\dag}_{-,#1}}

\begin{document}


\title{Hybrid light–matter excitations and spontaneous time-reversal symmetry breaking in two-dimensional Josephson Junctions}

\author{V. Varrica \orcidlink{0009-0002-2424-0044}}
 	\affiliation{Dipartimento di Fisica e Astronomia ``Ettore Majorana'', Universit\`a di Catania, Via S. Sofia 64, I-95123 Catania,~Italy}
 	\affiliation{INFN, Sez.~Catania, I-95123 Catania,~Italy}
\affiliation{Centro Siciliano di Fisica Nucleare e Struttura della Materia, Catania, Italy.}

\author{G. Falci \orcidlink{0000-0001-5842-2677}}%
\affiliation{Dipartimento di Fisica e Astronomia ``Ettore Majorana'', Universit\`a di Catania, Via S. Sofia 64, I-95123 Catania,~Italy}
\affiliation{INFN, Sez.~Catania, I-95123 Catania,~Italy}
%

\author{E. Paladino \orcidlink{0000-0002-9929-3768}}%
	\affiliation{Dipartimento di Fisica e Astronomia ``Ettore Majorana'', Universit\`a di Catania, Via S. Sofia 64, I-95123 Catania,~Italy}
	\affiliation{INFN, Sez.~Catania, I-95123 Catania,~Italy}
	%
\author{F.M.D. Pellegrino \orcidlink{0000-0001-5425-1292}}%
 	\affiliation{Dipartimento di Fisica e Astronomia ``Ettore Majorana'', Universit\`a di Catania, Via S. Sofia 64, I-95123 Catania,~Italy}
 	\affiliation{INFN, Sez.~Catania, I-95123 Catania,~Italy}
\affiliation{Centro Siciliano di Fisica Nucleare e Struttura della Materia, Catania, Italy.}

\date{\today}
\begin{abstract}
In the context of hybrid superconductor-semiconductor systems, Josephson junctions based on two-dimensional materials, such as graphene, offer promising opportunities because of their scalability and gate-tunable electronic properties.
In this work, we investigate the inductive coupling between a quantum LC resonator and a superconducting loop embedding a short, ballistic, planar Josephson junction, with the graphene-based case as a representative example. Within a mean-field formalism, we analyze how the properties of the global system depend on the light–matter interaction coupling, the Fermi level of the two-dimensional material, and temperature. 
Our findings reveal that the current-phase relation can show features indicative of spontaneous time-reversal symmetry breaking. Furthermore,  we determine the low-energy spectrum of collective hybridized light-matter excitations.
\end{abstract}
\maketitle

\section{Introduction}\label{sec:intro}
Hybrid systems that combine superconductors (S) and semiconductors (Sm) have been extensively studied in condensed matter physics due to their potential to host novel quantum phenomena, which could advance the development of scalable quantum technologies~\cite{clerk2020,aguado2020perspective,prada2020andreev,giannelli_ncimento_2022}.
Over the past decades, the integration of superconductivity, through the proximity effect, into Josephson junctions (JJs) based on low-dimensional semiconducting materials has paved the way for the achievement of coherent transport of Cooper pairs across normal conducting regions~\cite{nichele2020relating,valentini2024subgap}. Experimentally, the properties of these junctions have been explored within the circuit quantum electrodynamics (cQED) framework, which has ensured the manipulation of quantum degrees of freedom while also allowing microwave measurements by coupling the junctions with superconducting resonators~\cite{chidambaram2022,hinderling2023,elfeky2025}. 

A wide range of materials has been thoroughly investigated for their potential use as semiconducting components. These include InAs and InSb nanowires~\cite{kringhoj2021magnetic,matute2022signatures}, two-dimensional (2D) electron gases in III-V semiconductor heterostructures~\cite{kjaergaard2017hetero,marcus2023planar,strickland2024}, and  atomically thin materials such as graphene~\cite{li2018ballistic,pellegrino2020fnoise,dvir2021gjj}. Among them, 2D hybrid systems provide scalable platforms for future development of noise-protected qubits~\cite{fornieri2019,gyenis2021moving}, and their inherent tunability enables the realization of novel and complex devices~\cite{moehle2021,schiela2024}. Specifically, the graphene Josephson junction (GJJ) has emerged as a hybrid platform that supports highly transparent interfaces, allowing ballistic transport and constructive interference between Andreev reflections~, which coherently couple electron-hole pairs leading to the formation of Andreev bound states (ABSs)~\cite{li2016,park2018,schmidt2023tuning,beenakker2006,jois2023andrv,bretheau2017,park2024controllable}. Additionally, the transport properties of this platform have been improved by encapsulating graphene in hexagonal boron nitride, which is commonly used as a low-loss dielectric~\cite{wang2022hBN,ouaj2024hBN}. Recent advances have already demonstrated the compatibility of the GJJ with a variety of superconducting circuits, such as microwave cavities~\cite{schmidt2018} and gate-tunable transmon qubits~\cite{kroll2018,wang2019}. Furthermore, the low heat capacity of graphene combined with JJs nonlinearity has enabled the development of highly sensitive microwave bolometers~\cite{lee2020bolometer,kokkoniemi2020bolometer}, which can be integrated into gate-tunable parametric amplifiers that operate within the quantum-limited noise regime~\cite{butseraen2022gjjamp,sarkar2022gjjamp}. 

In JJs physics, a fundamental observable is the current-phase relation (CPR), which describes how the dissipationless supercurrent depends on the superconducting phase difference $\vphi$ across the junction~\cite{golubov2004}. In the case of GJJs, the CPR exhibits forward skewness compared to the traditional sinusoidal form, denoting the presence of higher harmonics~\cite{english2016current,messelot2024}, which are crucial for the design and application of superconducting devices~\cite{larsen2020parity, souto2022}. 
This characteristic depends on the microscopic composition of the junction, such as the number of conduction channels and their transmission properties~\cite{nanda2017current}. Moreover, recent experimental observations on quasi-2D InSb junctions have shown that skewness can be influenced by the gate voltage, the quality of the junction material, and the presence of spin-orbit coupling or magnetic fields~\cite{iorio2023,chieppa2025}.

Recent theoretical studies~\cite{park2020,metzger2021}, based on a perturbative approach, have provided a detailed description of the resonator frequency shifts observed experimentally in semiconducting weak links with only a few conduction channels. This approach works well when the two systems are sufficiently detuned. These works established a general framework for the readout of phase-biased superconducting weak links coupled to microwave resonators over a broad range of transition frequencies.

However, this description does not readily extend to systems with a large number of conduction channels. This is particularly relevant for planar platforms based on two-dimensional materials, such as graphene, which typically support many conduction channels~\cite{generalov2024,jung2025} and have been shown to host a large number of ABSs within the superconducting energy gap~\cite{wang2018}. In these systems, when the cavity frequency lies within the superconducting gap, multiple matter excitations can become exactly resonant with the electromagnetic mode, making a perturbative approach no longer valid.


%
In this work, motivated by the considerations above, we investigate the inductive interaction between a quantum LC circuit and a superconducting loop that hosts a 2D material-based JJ. We consider a short JJ, where the length of the junction, $L$,  is much smaller than the superconducting coherence length, $\xi$.  
When the width of the JJ, denoted as $W$, significantly exceeds the length $L$, resulting in a large geometric ratio $W/L \gg1$, the ABSs form a continuous spectrum within the superconducting energy gap $\deltaO$~\cite{pellegrino2022effect,vacante2024impurity}. This work addresses the scenario that will henceforth be designated as the wide short junction limit, with a particular focus on the specific instance of wide short GJJ~\cite{borzenets2016}.
Here, in the limit $W/L \gg1$, the energy splitting of several ABSs may become comparable to the characteristic energy of the quantum LC circuit, and we employ a mean-field approach that goes beyond the perturbative approach~\cite{park2020,metzger2021}.
%

When time-reversal symmetry is preserved, a conventional JJ exhibits a CPR that is both an odd function and $2\pi$-periodic~\cite{golubov2004}.
Therefore, this implies that no supercurrent flows when the superconducting phase difference is an integer multiple of $\pi$. We find that the inductive coupling between the loop hosting the GJJ and the superconducting resonator can generate a finite supercurrent at a superconducting phase difference $\varphi=\pi$, which represents a hallmark of time-reversal symmetry breaking (TRSB). 
This result presents a close analogy to the magnetostatic instability, also known as photon condensation~\cite{andolina_prb_2019,andolina_prb_2020,andolina_epjplus_2022,nataf_prl_2019,zueco_prl_2021,guerci_prl_2020,mazza_prb_2023,pellegrino_natcomm_2016,pellegrino_prb_2014,bamba_prl_2016,jaako_pra_2016,mercurio_prr_2024}.
%
This paper is structured as follows. Section~\ref{sec:model} provides details about the adopted model and introduces the theoretical framework that describes the interaction between the two circuits. In Section~\ref{sec:MF_theory}, we report analytical and numerical results based on the mean-field formalism at low temperatures, highlighting the influence of the light-matter interaction on the global system. Then, in Section~\ref{sec:polaritons}, within linear response, it is shown how hybridized light-matter excitations of the global system can be modulated due to the tunability provided by the platform. 
Finally, conclusions are drawn in Section~\ref{sec:conclusions}.

\section{Model}\label{sec:model}
\begin{figure}[t]
	\subfloat[]{%
        \begin{minipage}[t]{0.47\textwidth}
			\centering
			\includegraphics[width=0.70\textwidth]{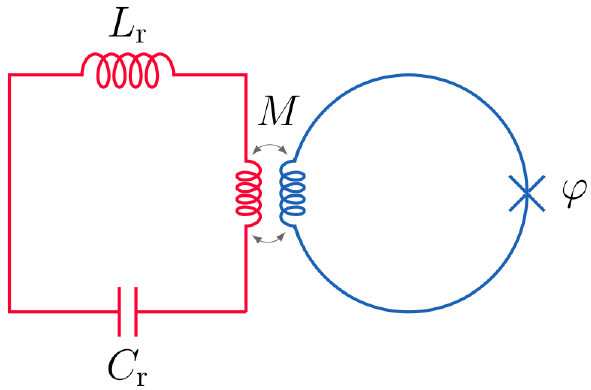}
			\label{fig:GJJ_LC_circuit_schematic}
        \end{minipage}
	}\hfill
	\subfloat[]{
        \begin{minipage}[t]{0.47\textwidth}
			\centering
			\includegraphics[width=0.95\textwidth]{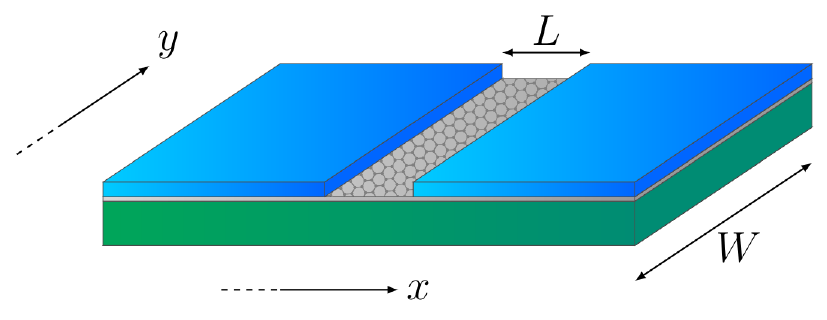}
			\label{fig:GJJ_schematic}
        \end{minipage}
	}
	\caption[]{(a) The resonant circuit is represented as a lumped-element LC resonator (red) with a capacitance $C_{\rm r}$ and an inductance $L_{\rm r}$. The LC circuit interacts inductively, through a mutual inductance $M$, with a loop containing a single short GJJ (blue). The superconducting phase difference across the GJJ is $\vphi$.  Here, the two coil-like elements connected by the small arrows provide a schematic illustration of the mutual inductive coupling. (b) A GJJ made by a monolayer graphene (grey) deposited on a substrate (green) and covered by two superconducting leads (blue). The uncovered grey region represents the graphene stripe in the normal phase. In this picture, $L$ represents the junction channel length along the $x$-direction (longitudinal) and $W$ is the width of the device along the $y$-direction (transverse).}
\end{figure}

We consider a superconducting loop with negligible self-inductance~\cite{metzger2021}, 
interrupted by a 2D material-based JJ. The loop is inductively coupled, via a mutual inductance 
$M$, to a superconducting resonator, as schematically illustrated by the two coupled coil-like elements in Fig.~\ref{fig:GJJ_LC_circuit_schematic}. The resonator is described as a lumped LC circuit characterized by capacitance $C_{\rm r}$ and inductance $L_{\rm r}$. Its quantum circuit Hamiltonian is given by 
\begin{equation}
	\hat{H}_{\rm r} = \hbar \omega_{\rm r}\left(\Ac\A+\frac{1}{2}\right)~,
	\label{eq:H_resonator}
\end{equation}
where the operator $\A$ ($\Ac$) destroys (creates) a photon in the resonator and $\omega_{\rm r}=1/\sqrt{L_{\rm r}C_{\rm r}}$ represents the characteristic resonator frequency~\cite{devoret2017}. Here, the resonator flux variable is identified with the operator $\hat{\Phi} = \Phi_{\rm{zpf}}(\A+\Ac)$~\cite{garcaripoll2022}, where $\Phi_{\rm{zpf}} = \sqrt{\hbar/(2C_{\rm r} \omega_{\rm r})}$ denotes the amplitude of zero-point flux fluctuations.

We describe the JJ as a short, wide junction that is homogeneous along the $y$-direction, as illustrated in Fig.~\ref{fig:GJJ_schematic}, and we model it within the Bogoliubov-de Gennes (BdG) approach~\cite{degennes1963,beenakker2008,reich2026}.
%
In the short junction regime, we focus on the eigenstates of the BdG Hamiltonian that have subgap eigenenergies, specifically $|E|<\Delta_0$, where $\Delta_0$ denotes the superconducting energy gap. These states, called ABSs, are solely responsible for carrying the Josephson equilibrium supercurrent~\cite{samuelson2000,levchenko2006}.
The Andreev continuum, which comprises eigenstates with energies above the gap, $|E|>\Delta_0$, is not considered.
In the wide limit, the electron system of a short junction can be described as a macroscopic amount of ABSs that form energy level pairs within the superconducting energy gap~\cite{titov2006, pellegrino2022effect, vacante2024impurity}. 
We introduce the ABS fermionic annihilation operator $\hat{\gamma}_{j,k}$ labeled by $j$, which denotes if the ABS has an eigenenergy above ($j=+$) or below ($j=-$) the Fermi level, and the indices $k$ label the propagating channels within the normal stripe~\cite{vacante2024impurity}.
In the case of a GJJ, $k$ is replaced by the composite index $(\zeta, k)$, where $k$ is the usual transverse component of the wavevector, which we express in units of $1/L$, and  $\zeta = \pm$  denotes the valley index. Given that the valley degree of freedom is not active, it merely contributes a twofold degeneracy in the observables. 
Thus, when analyzing the GJJ, we omit the valley index and perform the calculations for a single valley, restoring the valley degeneracy at the end by including a factor $n_{\rm v}=2$.
In the short junction limit, when $L\ll \xi \sim \hvf/\deltaO$ where $v_{\rm F}$ is the Fermi velocity, only one pair of ABSs is expected for each conduction channel~\cite{beenakker1991}. A pair of ABSs is characterized by the energies $\pm \epsilon(k, \vphi)$, where
\begin{equation}
	\epsilon(k, \vphi)=\deltaO\sqrt{1-\tau(k)\Msinq{\vphi/2}}~,
	\label{eq:ABSs_energy}
\end{equation}
where $\vphi$ is the superconductive phase difference through the junction and $\tau(k)$ is the normal phase transmission probability~\cite{beenakker1991}.
For GJJ, within the Dirac-BdG approach, the transmission probability~\cite{titov2006} is given by 
\begin{equation}\label{eq:tauG}
\tau(k) = \frac{\kf^2- k^2}{\kf^{2} - k^2\cos^2{(\sqrt{\kf^{2} -k^2})}}~, 
\end{equation}
where $\kf = \abs{\mu_0}L/\hvf$ is the Fermi wavenumber in units of $1/L$, $\mu_0$ is the Fermi level in the graphene stripe, and $v_{\rm F}\approx c/300$. The resulting low-energy Hamiltonian can be expressed on the Andreev basis as
\begin{equation}
	\hat{H}_{\rm A} =  \mathcal{N} \int_{-\infty}^{+\infty} \frac{dk}{2\pi} \epsilon(k,\vphi)\sigmaz_{k}~,
	\label{eq:H_Andreev}
\end{equation}
where and $\sigmaz_{k}={\hat{\gamma}^{\dag}}_{+,k}\hat{\gamma}^{\mathstrut}_{+,k} -   \hat{\gamma}^{\dag}_{-,k}\hat{\gamma}^{\mathstrut}_{-,k}$ denotes the diagonal Pauli operator acting on the even-parity $2\times2$ ABSs subspace, and $\mathcal{N} =  W/L\gg1$. 

The mutual inductive coupling between two circuits with self-inductances $L_1$ and $L_2$ is commonly expressed as $M = \kappa_M \sqrt{L_1 L_2}$, where the coupling factor $\kappa_M$ satisfies $0 \leq \kappa_M \leq 1$ to ensure that the magnetic energy remains positive~\cite{alexander2021fundamentals}.
The magnetic flux $\phi_1$ in the first circuit is related to the current $I_1$ flowing in that circuit by $\phi_1 = L_1 I_1$. Due to mutual inductance, the resulting magnetic flux in the second circuit is given by $\phi_{21} = M I_1$. Consequently, these two fluxes can be connected through the relation $\phi_{21} = \frac{M}{L_1} \phi_1=\kappa_M\sqrt{\frac{L_2}{L_1}}\phi_1$.

In our description, the first circuit is the LC-resonator, so we identify $L_1 \equiv L_{\rm r}$, whereas $L_2$ represents the effective inductance of $\mathcal{N}$ single-channel weak links connected in parallel, which lie in the JJ. Because there are $\mathcal{N}$ parallel channels, the corresponding self-inductance scales as $L_2 \sim 1/\mathcal{N}$. Thus, by expressing the mutual inductance in terms of the coupling coefficient $\kappa_M$ and the self-inductances, one finds that it scales as $M \sim 1/\sqrt{\mathcal{N}}$.
Using the discussion above, the mutual inductance induces in the JJ a flux given by $\frac{M}{L_{\rm r}} \hat{\Phi}$. Consequently, the gauge-invariant phase across the JJ, with superconducting phase difference $\varphi$, can be written as $\varphi+\hat{\varphi}_M$, where
$\hat{\varphi}_M = \frac{M}{\phi_0 L_{\rm r}}\hat{\Phi} = \frac{M \Phi_{\rm{zpf}}}{\phi_0 L_{\rm r}} (\A+\Ac)$ denotes the induced flux divided by the reduced flux quantum $\phi_0 = \hbar/(2e)$.
In line with Refs.~\cite{park2020,metzger2021,zazunov2003,bretheau2013}, treating the term $\hat{\varphi}_M$ as a small quantity, the effective low-energy Hamiltonian governing the interaction between the ABSs and the resonator is obtained by expanding the Andreev Hamiltonian $\hat{H}_{\rm A}$ in powers of $\hat{\varphi}_M$ and keeping terms up to second order. This gives $\hat{H}_{\rm int}=\hat{\varphi}_M \partial_\varphi \hat{H}_{\rm A}+\frac{1}{2}\hat{\varphi}_M^2 \partial^2_\varphi \hat{H}_{\rm A}$.


The total Hamiltonian that describes the two circuits and their inductive interaction is given by
\begin{equation}
	\hat{H}_{\rm} = \hat{H}_r + \hat{H}_{\rm A} + \frac{g}{\sqrt{\mathcal{N}}} (\A + \Ac)\df\hat{H}_{\rm A} + \frac{g^2}{2\mathcal{N}} (\A + \Ac)^2 \dqf\hat{H}_{\rm A}~,
	\label{eq:H_tot}
\end{equation}
Here, $g = M\Phi_{\rm{zpf}}\sqrt{\mathcal{N}}/(L_{\rm r}\phi_0)$ denotes a dimensionless coupling constant that characterizes the magnitude of the inductive interaction.
We observe that because the mutual inductance $M$ scales as $\sim 1/\sqrt{\mathcal{N}}$, the coupling $g$ does not depend on $\mathcal{N}$ in the limit $\mathcal{N}\gg1$.

In Eq.~\eqref{eq:H_Andreev}, the Hamiltonian $\hat{H}_{\rm A}$ is expressed on the Andreev basis, whose eigenstates depend on the superconducting phase difference $\vphi$. Consequently, the derivatives of the Hamiltonian $\hat{H}_{\rm A}$ with respect to $\vphi$ require differentiating both the eigenenergies and the basis states. The resulting expressions are
\begin{widetext}
\begin{subequations}
	\begin{align}
		\df\hat{H}_{\rm A} &= \mathcal{N}\int_{-\infty}^{+\infty} \frac{dk}{2\pi} \df\epsilon(k,\vphi)\bigg[\sigmaz_{k} - \sqrt{1-\tau(k)}\Mtan{\frac{\vphi}{2}}\sigmax_{k}\bigg]~, \label{eq:Current_term_andreev_continuum}\\
		  \dqf\hat{H}_{\rm A} & = \mathcal{N}\int_{-\infty}^{+\infty}  \frac{dk}{2\pi}\df\epsilon(k,\vphi) \bigg[\frac{\tau(k) +(2-\tau(k)) \Mcos{\vphi}}{2\Msin{\vphi}} \sigmaz_{k} - \sqrt{1-\tau(k)}\sigmax_{k}\bigg]~,\label{eq:Diamagn_term_andreev_continuum}
	\end{align}
\label{eq:high_order_term_andreev_continuum}
\end{subequations}
\end{widetext}
where $\sigmax_{k}={\hat{\gamma}^{\dag}}_{+,k}\hat{\gamma}^{}_{-,k} +   \hat{\gamma}^{\dag}_{-,k}\hat{\gamma}_{+,k}$ represents the off-diagonal Pauli operator in the Andreev basis. The full derivation of Eq.~\eqref{eq:high_order_term_andreev_continuum} is provided in Appendix~\ref{app:Andreev_basis}.

Here, $\hat{I}_{\rm A}=\frac{1}{\phi_0} \df\hat{H}_{\rm A}$ corresponds to the Andreev current operator~\cite{zazunov2003}. Its diagonal term gives the equilibrium supercurrent carried by the ABSs, while the off-diagonal term involves single ABSs transitions and is responsible for current fluctuations~\cite{pellegrino2022effect}.
Whereas Eq.~\eqref{eq:Diamagn_term_andreev_continuum} is associated with the inverse inductance operator, $\hat{\cal L}_{\rm A}^{-1}=\frac{1}{\phi_0^2} \dqf\hat{H}_{\rm A}$, that, evaluated at equilibrium, gives the reciprocal of the characteristic Josephson kinetic inductance~\cite{bretheau2013,baumgartner2021induct}.

\section{Mean-Field theory}\label{sec:MF_theory}
\begin{figure*}[t]
	\subfloat[]{
		\begin{minipage}[t]{0.48\textwidth}
			\centering
			\includegraphics[height=0.225\textheight]{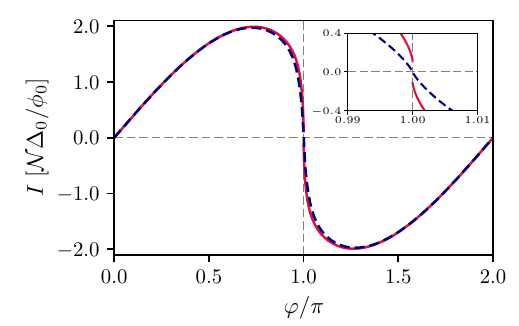}
			\label{fig:I_vs_phi_g_comparison}
		\end{minipage}
	}
	\subfloat[]{
		\begin{minipage}[t]{0.48\textwidth}
			\centering
			\includegraphics[height=0.225\textheight]{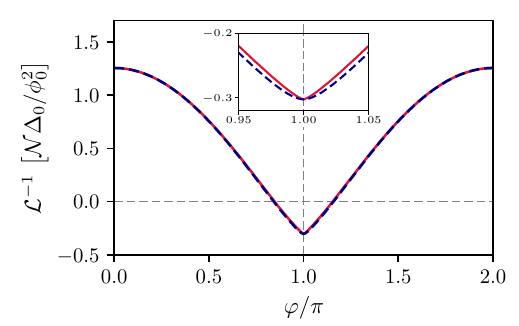}
			\label{fig:L_vs_phi_g_comparison}
		\end{minipage}	
	}
	\caption[]{%
    CPR and reciprocal inductance of the GJJ for coupling constant $g = 0.1$ (red solid line) and  $g = 0$ (blue dashed line), shown respectively in (a) and (b). The quantities are expressed in units of  $\mathcal{N}\Delta_0/\phi_0$ and $\mathcal{N}\Delta_0/\phi_0^{2}$. In the inset of (a), two finite supercurrents of opposite sign appear at $\varphi=\pi$ when the coupling constant $g$ is switched on. In the inset of (b), the coupling produces only a small renormalization near $\varphi=\pi$. All results are obtained at zero temperature. Other parameters are  $\hbar\omega_{\rm r}=0.6\Delta_0$ and $\mu_0 = 10\hbar v_{\rm F}/L$.
    }
	\label{fig:MF_results_T0}
\end{figure*}

In this Section, we investigate the main equilibrium characteristics of the global system using a mean-field approach.
The many-body Hamiltonian in Eq.~\eqref{eq:H_tot} presents both light ($\A$) and matter ($\hat{\gamma}_{j,k}$) degrees of freedom combined by an interaction that has a non-quadratic structure. 
Diagonalization of this many-body Hamiltonian is impractical. Therefore, employing a mean-field approach is a feasible strategy~\cite{bruus2004many}.
According to this scheme, the light and matter degrees of freedom are effectively decoupled. Each one interacts with an effective mean field whose value is determined self-consistently at equilibrium.
The full derivation of the mean-field Hamiltonian is reported in App.~\ref{app:Self-consistent problem}.
%
We restrict our analysis to the low-temperature regime $\kBT/\deltaO\leq10^{-2}$,
where the superconducting features of the leads are not significantly affected by thermal effects. In particular, we neglect the temperature dependence of the superconducting energy gap.
According to the mean-field scheme, the Hamiltonian is represented as follows
\begin{subequations}
	\begin{align}
		\hat{H}_{\rm{MF}} &= \hat{H}_{\rm B}+\hat{H}_{\rm F}+{\cal N}(2\alpha {\cal P}-4\alpha^2 {\cal D})~,\label{eq:H_MF_complete}\\
		\hat{H}_{\rm F} &= \hat{H}_{\rm A} - 2 g\phi_0\alpha \hat{I}_{\rm A}
        +2\left( g\phi_0\alpha \right)^2 \hat{\cal L}_{\rm A}^{-1}
        ~,\label{eq:H_MF_fermionic}        \\
		\hat{H}_{\rm B} &= \hat{H}_{\rm r} + (\A + \Ac)\sqrt{\mathcal{N}}\mathcal{P} + (\A + \Ac)^{2}\mathcal{D}~,\label{eq:H_MF_bosonic}
	\end{align}	
	\label{eq:H_MF_terms}
\end{subequations}
where the mean fields are expressed as 
\begin{subequations}
	\label{eq:MF_terms}
	\begin{align}
		\mathcal{P}&=\frac{g \phi_0 n_{\rm v}}{\mathcal{N}} \expval*{\hat{I}_{A}}_{\rm MF}~,\label{eq:MF_P}\\
		%
		%
		\mathcal{D}&=\frac{(g \phi_0)^2 n_{\rm v}}{2\mathcal{N}}\expval*{\hat{\cal L}^{-1}_{A}}_{\rm MF}~,\label{eq:MF_D}\\
		%
		\alpha&=-\frac{1}{2\sqrt{\cal N}}\expval*{\A+\Ac}_{\rm MF} 
        \label{eq:MF_alpha}~,
	\end{align}	
\end{subequations}
they are real-valued quantities and $n_{\rm v}$ represents the potential valley degeneracy factor.
Here, we denote $\expval{\cdot}_{\rm MF}= \Tr( \hat{\rho}_{\rm MF} \cdot )/{\cal Z}_{\rm MF}$, where $\hat{\rho}_{\rm MF}=e^{- \beta \hat{H}_{\rm MF} }$ is the mean-field thermal density matrix,
$\beta$ is the reciprocal of the thermal energy $1/(k_{\rm B} T)$,  ${\cal Z}_{\rm MF}=  \Tr( \hat{\rho}_{\rm MF} ) $ is the mean-field partition function, and the trace $\Tr(\cdot)$ is taken over both the fermionic ABSs sector and the bosonic degrees of freedom.
Specifically, the mean-field $\mathcal{P}$ is related to the supercurrent flowing in the loop  $I=n_{\rm v}\expval*{\hat{I}_{\rm A}}_{\rm MF}$,
and the mean-field $\mathcal{D}$ pertains to the inverse of the non-linear inductance of the Josephson element ${\cal L}^{-1}=n_{\rm v}\expval*{\hat{\cal L}^{-1}_{\rm A}}_{\rm MF}$.
%
%
Finally, $\alpha$ represents a shift induced in the expectation value of the resonator coordinate, which is zero when evaluated in the bare ground state of the LC quantum oscillator. 
According to this scheme, diagonalizing the quadratic bosonic Hamiltonian $\hat{H}_{\rm B}$ in Eq.~\eqref{eq:H_MF_bosonic}, as detailed in
Appendix~\ref{app:Self-consistent problem}, and subsequently evaluating the equilibrium expectation value of the coordinate operator $\expval*{\hat{a}+\hat{a}^\dagger}_{\rm MF}$, leads to 
\begin{equation}\label{eq:MF_alpha-PD}
        \alpha=\frac{\mathcal{P}}{\hbar\omega_{\rm r}  + 4\mathcal{D}}~,
\end{equation}
which expresses $\alpha$ in terms of ${\cal P}$ and ${\cal D}$.

The matter sector is described by the single-particle fermionic Hamiltonian $\hat{H}_{\mathrm{F}}$ introduced in Eq.~\eqref{eq:H_MF_fermionic}, where $\alpha$ plays the role of an external field. 
By diagonalizing $\hat{H}_{\mathrm{F}}$, one can determine the equilibrium values of the quantities $I$ and $\mathcal{L}^{-1}$. These values are directly proportional to the mean-fields $\mathcal{P}$ and $\mathcal{D}$, respectively, and exhibit a nonlinear dependence on $\alpha$.
By combining the dependence of $\alpha$ on $\mathcal{P}$ and $\mathcal{D}$ with Eq.~\eqref{eq:MF_alpha-PD}, the self-consistent approach can be expressed as a set of nonlinear equations for $\mathcal{P}$ and $\mathcal{D}$, or equivalently as a nonlinear equation for $\alpha$.
%
In App.~\ref{app:Self-consistent problem} we show explicitly the self-consistent approach at a generic temperature, which is solved numerically.

%
%
Here, firstly, we discuss the case at zero temperature.
According to the structure of the mean-field Hamiltonian, the ground state can be expressed as a product state
\begin{subequations}
\begin{align}\label{eq:MFGS}
\ket{\Psi_0}&=\ket{F_0}\otimes\ket{B_0}~, \\
\ket{F_0}&=\prod_{ k} [\sin(\theta_k/2)\hat{\gamma}_{+,k}^\dagger \hat{\gamma}_{-,k}+\cos(\theta_k/2) ]\ket{\emptyset}~, \label{eq:MF-matGS}
\end{align}
\end{subequations}
where $\ket{\emptyset}$ refers to the state $\prod_{k} \hat{\gamma}^\dagger_{-,k}\ket{\rm vac}$, which is fully populated by quasiparticles that occupy the lower ABSs. Consequently, $\ket{F_0}$ is composed of a product of linear combinations involving pairs of ABSs specified by the index $k$.
Here, the angle $\theta_k$ depends on the Andreev Hamiltonian $\hat{H}_{\rm A}$ and its derivatives with respect to $\varphi$, the dimensionless parameter $g$, and the mean-field $\alpha$, while $\ket{B_0}$ represents the ground-state of the quadratic bosonic Hamiltonian $\hat{H}_{\rm B}$.
By turning off the coupling constant $g$, the ground state takes the form in which 
all $\theta_k=0$, and $\ket{B_0}$ corresponds to the vacuum state $\ket{0}$ of the photon annihilation operator, $\hat{a} \ket{0}=0$.

At zero temperature, the self-consistent procedure looks for the global minimum of the energy density functional, defined as
\begin{equation}\label{eq:calE}
\begin{aligned}
{\cal E}(\vphi)=
\frac{\bra{\Psi_0} \hat{H} \ket{\Psi_0}}{\cal N}=\hbar \omega_{\rm r}\alpha^2- n_{\rm v} \int_{-\infty}^\infty \frac{dk}{2\pi}E(k,\vphi)~,
\end{aligned}
\end{equation}
where $n_{\rm v}$ represents the potential valley degeneracy,
\begin{subequations}\label{eq:mf-Ek}
	\begin{align}
         E(k,\vphi)&=\sqrt{\cdx{k}^2+\cdz{k}^2}~,\\
		\cdx{k} &=  2g\alpha\df\epsilon(k,\vphi)\sqrt{1-\tau(k)}\bigg(\Mtan{\frac{\vphi}{2}} - g\alpha \bigg)~,\\
		\cdz{k} &= \epsilon(k,\vphi) - 2g\alpha\df\epsilon(k,\vphi)\\
        &\times\bigg(1 -  g\alpha\frac{\tau(k) +(2-\tau(k)) \Mcos{\vphi}}{2\Msin{\vphi}}\bigg)~.\nonumber
		%
	\end{align}
\end{subequations}
The functional ${\cal E}$ remains finite in the wide junction limit ${\cal N}\to \infty$, as it represents the ground energy of the mean-field Hamiltonian divided by $\cal N$.
Above, we have expressed the self-consistent problem in terms of $\alpha$.
Moreover, within the self-consistent approach, the dependence on $k$ enters in all quantities by means of $\tau(k)$, as it is clearly visible in Eq.~\eqref{eq:mf-Ek}. For this reason, it is useful to introduce a density of state (DOS) resolved in normal-phase transmission probability~\cite{cuevas2006},
\begin{equation}
	\rho(\tau) = \int_{-\infty}^{+\infty}\frac{dk}{2\pi} \delta(\tau(k)-\tau)~,
	\label{eq:DOS_tau}
\end{equation}
where $\delta(x)$ denotes the Dirac's delta function.
Consequently, integrals expressed as $\int_{-\infty}^{+\infty}\frac{dk}{2\pi} J[\tau(k)]$, where the integrand consists of a generic function $J$ composed with $\tau(k)$, can be represented equivalently as $\int_0^1 d \tau \rho(\tau)J(\tau)$.
As an illustrative example, we consider the GJJ case and discuss the results obtained.
Fig.~\ref{fig:I_vs_phi_g_comparison} displays the CPR of the GJJ at zero temperature for a finite value of the coupling constant $g =0.1$ (red solid line), compared with the case of an isolated GJJ, $g = 0$ (blue dashed line), where the expression for the supercurrent takes the usual form $I= -\frac{\mathcal{N} n_{\rm v}}{\phi_0}\int_{-\infty}^{+\infty}  \frac{dk}{2\pi}\df\epsilon(k,\vphi) $, with $n_{\rm v}=2$~\cite{titov2006}.
Here, we fix the Fermi level and bare cavity energy at the generic values $\mu_0=10 \hbar v_{\rm F}/L$ and  $\hbar \omega_{\rm r}=0.6 \Delta_0$, respectively. 
The supercurrent $I$ is shown in units of ${\cal N} \Delta_0 /\phi_0$, which, for  $g\neq0$, corresponds to the solution of the self-consistent problem ${\cal P}$ expressed in units of $g \Delta_0$. 
Here, the two curves that show the CPR are essentially coincident, except for the region around $\vphi =\pi$, which is more clearly illustrated in the inset.
In particular, at $g=0.1$, the CPR exhibits a jump discontinuity at $\vphi = \pi$ characterized by two finite supercurrent values in opposite directions occurring at $\varphi = \pi^-$ and $\varphi = \pi^+$. In contrast, with $g=0$, the CPR behaves continuously, indicating the absence of supercurrent at $\vphi = \pi$.
Similarly, Fig.~\ref{fig:L_vs_phi_g_comparison}  shows the inverse of the GJJ inductance as a function of the phase difference $\vphi$, at zero temperature and for a finite value of the coupling constant $g =0.1$ (red solid line), and it is compared to the case of an isolated GJJ, $g = 0$ (blue dashed line). 
Here, we observe that the deviation between these curves is minimal throughout the range of $\vphi$.

In conventional JJs, the supercurrent cannot flow at $\vphi=n \pi$ (where $n$ is any integer) due to time-reversal symmetry~\cite{golubov2004}. 
On the other side, JJs made with weak ferromagnetic links or with materials that host spin-orbit coupling in the presence of a magnetic field~\cite{trimble2021jjsymmetry,diez2023jjsymmetry}, sustain supercurrent at $\vphi=\pi$, and this is a signature of TRSB. 
The lack of supercurrent at $\vphi=n \pi$  remains valid for a GJJ isolated from an LC quantum harmonic oscillator.
Furthermore, the microscopic Dirac-BdG Hamiltonian, which describes the complete electron system of a GJJ~\cite{titov2006}, and from which the Andreev Hamiltonian is derived \cite{pellegrino2022effect, vacante2024impurity}, retains explicit time-reversal symmetry when $\vphi$ is set at $n \pi$.
In our global system, although the total Hamiltonian is invariant under time reversal, the interaction between the GJJ and an LC quantum harmonic oscillator leads to the emergence of a non-zero supercurrent at $\varphi=\pi$.  This effect represents a hallmark of spontaneous TRSB arising from inductive coupling.
Considering the GJJ isolated from the LC quantum harmonic oscillator, the expectation value of the current operator $\hat{I}_{\rm A}$ for generic ABSs $\hat{\gamma}_{j, k}^\dagger\ket{\rm vac}$ at $\varphi = \pi^+$ and $\varphi = \pi^-$ is zero, except for ABSs associated with total transmission.
In these specific cases, for values of $k$ such that transmission $\tau(k) = 1$, the corresponding pairs of degenerate ABSs have zero-energy splitting~\cite{fatemi2025}.  When the GJJ is isolated from the LC quantum harmonic oscillator, its ground state hosts pairs of counter-propagating supercurrents of equal magnitude, sustained by the degenerate ABSs associated with total transmission. This zero-net-current configuration becomes unstable once the coupling between the GJJ and the LC quantum harmonic oscillator is turned on ($g \neq 0$), the interaction between the two circuits then breaks the current balance, producing a finite supercurrent at $\varphi = \pi$.
%

Beyond the intuitive picture presented above, we now provide a more formal and quantitative description of the phenomenology discussed, which is valid for any short and wide 2D material-based JJ coupled with an LC quantum harmonic oscillator.
By calculating the expectation value of the current operator, see Eq.~\eqref{eq:Current_term_andreev_continuum}, on the ground-state of the matter sector $\ket{F_0}$, shown in Eq.~\eqref{eq:MF-matGS}, for $\varphi=\pi$, one finds
$\bra{F_0} \hat{I}_{\rm A}\ket{F_0}=\frac{\mathcal{N} \Delta_0}{\phi_0} \int_{-\infty}^{+\infty} \frac{dk}{2\pi} \tau(k)   \sin(\theta_k)  $. 
This result indicates that, at $\varphi = \pi$, the supercurrent is nonzero if there is a finite range of $k$ such that $\theta_k \neq n\pi$ ($n$ is an integer), namely, if the ground state is populated by quasiparticles that occupy states which are expressed as linear combinations of the upper ($j=+$) and lower ($j=-$) ABSs.
In App.~\ref{app:Self-consistent problem}, one finds the dependence of $\theta_{k}$ on the mean fields introduced in Eqs.~\eqref{eq:MF_terms}.
For small $\alpha$, such that $|\alpha|\ll 1$, one has 
$\sin(\theta_k)\approx g \alpha \tau(k)/\sqrt{1-\tau(k)}$ that means that a finite value of $\alpha$ is enough to generate a supercurrent at $\vphi=\pi$. 
Moreover, for small values of $|g \alpha|$, using Eq.~\eqref{eq:mf-Ek}, one finds
\begin{equation}\label{eq:Ek_pi}
E(k,\pi)\approx\Delta_0 \sqrt{1-[1-(g\alpha)^2]\tau(k)}~,
\end{equation}
by expanding it up to $\order{(g\alpha)^2}$, the energy density functional, at $\vphi=\pi$, can be written as
\begin{equation}\label{eq:energy_exp}
\begin{aligned}
\frac{{\cal E}(\pi)}{\deltaO} &\approx \varepsilon_0 + \left[\frac{\hbar\omega_{\rm r}}{\deltaO}- \frac{n_{\rm v}g^2}{2}\int_{0}^1 d \tau  \frac{\rho(\tau)\tau}{\sqrt{1-\tau}}\right]\alpha^2,
\end{aligned}
\end{equation}
where $\varepsilon_0=- n_{\rm v}\int_{0}^1 d \tau \rho(\tau)\sqrt{1-\tau}$.
The expression above defines an instability condition, which is fulfilled when the quantity within the square brackets becomes non-positive. In this case, the ground-state energy no longer exhibits a minimum at $\alpha = 0$. Since the energy density functional is bounded from below, it must therefore possess at least one minimum at a finite value of $\alpha$.
In particular, we define the critical coupling constant $g_{\rm c}$ as
\begin{equation}\label{eq:gc}
g_{\rm c}= \left[ \frac{ n_{\rm v} \Delta_0}{2\omega_{\rm r}} \int_{0}^1 d \tau \frac{\rho(\tau) \tau}{\sqrt{1-\tau}} \right]^{-1/2}~,
\end{equation}
such that for $|g| \ge g_{\rm c}$ the mean-field ground-state energy is minimized at a finite value of $\alpha$, which, according to the discussion above, corresponds to a finite supercurrent.

We recall that the definition of $g_{\rm c}$ and the discussion of the instability are not specific to the case of a GJJ, since we have not yet specified the form of $\tau(k)$. 
For the normal state of the graphene electron gas~\cite{titov2006}, there are propagating channels with total transmission,  $\tau(k)=1$. 
In particular, total transmission occurs for $k_0=0$ (Klein tunneling~\cite{katsnelson_book})  and for  $ k_i=\sgn(i) \sqrt{[\mu_0 L/(\hbar v_{\rm F})]^2-(i \pi)^2}$ (stationary-wave condition), where 
$i=\pm1,\ldots,\pm N$ ($N=\lfloor \mu_0 L/(\pi\hbar v_{\rm F})\rfloor$, where $\lfloor \cdot\rfloor$ denotes the integer part). 
Since $\tau(k)$ is a bounded function, $0\leq\tau(k)\leq1$, the propagating channels with total transmission correspond to a global maximum of the transmission probability. Consequently, the density of states $\rho(\tau)$ exhibits an integrable square-root divergence at $\tau=1$~\cite{pellegrino2022effect,vacante2024impurity}. 
It is useful to isolate this divergence and express the DOS as
$\rho(\tau)=\rho_0/\sqrt{1-\tau}+\rho_1(\tau)$.
The coefficient $\rho_0$ quantifies the weight of the transparent modes in the DOS and depends on the value of the Fermi level $\mu_0$. 
In particular, $\rho_0$ is determined by the curvatures of the transmission probability at the global maxima, $k_i$ with $i=-N,\ldots,N$, see the details in App.~\ref{app:DOS_approximation}. 
Conversely, as $\tau\to1$, $\rho_1(\tau)$ exhibits regular behavior, and its form is determined by $\mu_0$.
By introducing the expression for $\rho(\tau)$ of a GJJ, discussed above, into Eq.~\eqref{eq:gc}, the integral develops a logarithmic divergence, yielding
$g_c \propto  1/\sqrt{\lim_{\tau \to 1^-}| \ln(1-\tau)|}$,
which implies that the critical coupling constant is infinitesimal.
%
Moreover,  combining Eq.~\eqref{eq:Ek_pi} with the decomposition $\rho(\tau)=\rho_0/\sqrt{1-\tau}+\rho_1(\tau)$, for GJJ, in Eq.~\eqref{eq:calE}, one has an explicit form for the energy density functional 
\begin{equation}
\begin{aligned}
\frac{{\cal E}(\pi)}{\deltaO } &\approx \varepsilon_0 +\bigg\{\frac{\hbar\omega_{\rm r}}{\deltaO}
+\frac{n_{\rm v}\rho_0 g^2}{\sqrt{1-(g\alpha)^2}}\ln \bigg[ \frac{|g \alpha|}{1+\sqrt{1-(g\alpha)^2}} \bigg]  \\
&-n_{\rm v}g^2\int_{0}^1 d \tau  
 \frac{\rho_1(\tau) \tau}{\sqrt{1-[1-(g\alpha)^2]\tau}+\sqrt{1-\tau}} \bigg\}\alpha^2.
\end{aligned}
\end{equation}
\begin{figure}[t]
    \centering
    \includegraphics[width=0.475\textwidth]{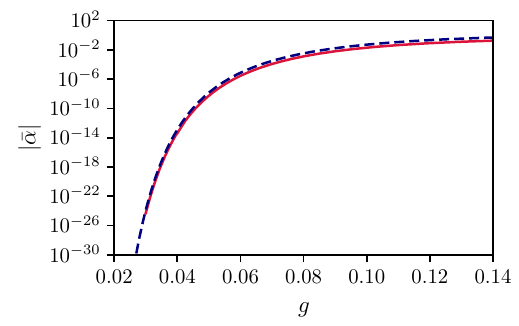}
	\caption[]{%
    Modulus of the photonic mean-field $\bar{\alpha}$ minimizing the energy-density functional as a function of the 
    coupling constant $g$ at zero temperature. The red line shows the value of $\bar{\alpha}$ obtained 
    from the full numerical solution of the self-consistent approach, while the blue dashed line 
    corresponds to Eq.~\eqref{eq:baralpha}, with $\kappa=0$.  Because of the logarithmic 
    singularity of the energy-density functional in the variable $g\alpha$ as $g\alpha\to0$, the 
    numerical search is terminated at $g \lesssim 0.03$, where $|\bar{\alpha}|\sim10^{-25}$, beyond which resolution limitations prevent a 
    stable solution. Other parameters are $\hbar\omega_{\mathrm r} = 0.6\Delta_0$ and 
    $\mu_0 = 10\hbar v_{\mathrm F}/L$.
    }
    \label{fig:alpha_VS_g}
\end{figure}

\begin{figure*}[t]
	\subfloat[]{
		\begin{minipage}[t]{0.48\textwidth}
			\centering
			\includegraphics[height=0.225\textheight]{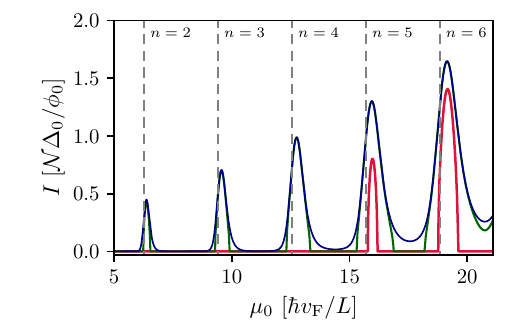}
            \label{fig:I_pi_vs_mu_T_comparison}
		\end{minipage}
	}
	\subfloat[]{
		\begin{minipage}[t]{0.48\textwidth}
			\centering
			\includegraphics[height=0.221\textheight]{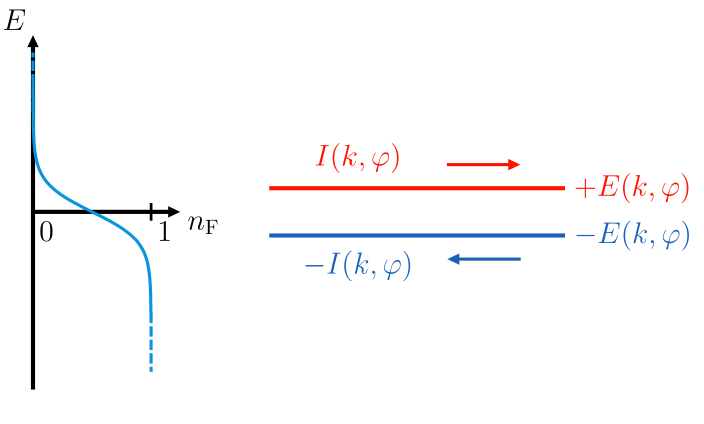}
			\label{fig:MF_occupancy}
		\end{minipage}	
	}
	\caption[]{%
    (a) Supercurrent $I$ at $\varphi = \pi^{-}$, in units of $\mathcal{N}\Delta_0/\phi_{0}$, as a function of the Fermi level $\mu_{0}$, expressed in units of $\hbar v_{\mathrm{F}}/L$, for three temperatures: $k_{\mathrm{B}}T/\Delta_0 = 0$ (blue solid line), $0.002$ (green solid line), and $0.01$ (red solid line). Vertical dashed lines mark values of $\mu_{0} L/(\hbar v_{\mathrm{F}})$ equal to  $n \pi$ (where $n$ is an integer). The remaining parameters are $\hbar\omega_{\mathrm{r}} = 0.6 \Delta_0$ and $g = 0.1$.
    (b)  Schematic representation of a pair of mean-field quasiparticle energy levels, $\pm E(k,\vphi)$, corresponding to a high-transmission channel ($\tau(k)\lesssim 1$) for a generic superconducting phase difference $\vphi$. The red (blue) level refers to the upper (lower) quasiparticle state. While the horizontal arrows indicate the direction of the corresponding supercurrent contribution, $\pm I(k,\vphi)=\expval{\hat{I}_{\rm A}}{\psi_{\pm,k}}$.  The Fermi-Dirac distribution at low temperature indicates that, as temperature increases, the upper (red) quasiparticle state becomes thermally populated.
    }
	\label{fig:fig:I_pi_vs_mu_T}
\end{figure*}

Focusing on the limit $(g\alpha)^2 \ll1 $, one obtains
\begin{equation}
\frac{{\cal E}(\pi)}{\deltaO } \approx \varepsilon_0 +\left\{\frac{\hbar \omega_{\rm r}}{\deltaO}
+n_{\rm v}\rho_0 g^2 \left[\ln\left(\frac{g |\alpha|}{2}\right) -\kappa_1\right] \right\}\alpha^2~,
\label{eq:MF_energy_density_functional}
\end{equation}
where $\kappa_1=(2 \rho_0)^{-1}\int_0^1 \rho_1(\tau) \tau/\sqrt{1-\tau}$. Accordingly, the energy density functional exhibits a minimum at $\bar{\alpha}$ such that
\begin{equation}\label{eq:baralpha}
|\bar{\alpha}|=\frac{1}{g}\exp\bigg[-\bigg(\frac{1}{g^2}\frac{\hbar \omega_{\rm r}}{n_{\rm v} \rho_0 \Delta_0}-\kappa\bigg)\bigg]~,
\end{equation}
where $\kappa=\kappa_1-\frac{1}{2}+\ln(2)$ is a function of $\mu_0$ through $\kappa_1$.
We note that employing  Eq.~\eqref{eq:mf-Ek}, instead of the approximate expression in Eq.~\eqref{eq:Ek_pi}, would result in a renormalization of the coefficient $\kappa$.
The finding in Eq.~\eqref{eq:baralpha} demonstrates that for any given coupling constant $g$, the energy density functional exhibits two degenerate global minimum points at $\pm |\bar{\alpha}|$, which is directly associated with the observation in Fig.~\ref{fig:I_vs_phi_g_comparison}, where the supercurrent at $\vphi=\pi^-$ and $\vphi=\pi^+$ takes on two opposite values with the same magnitude.
Fig.~\ref{fig:alpha_VS_g} shows the value $\abs{\bar{\alpha}}$ that minimizes the energy functional density ${\cal E}(\pi)$ as a function of the coupling constant $g$.
This is determined by numerically solving the self-consistent problem at zero temperature (red solid line). For comparison, the closed-form expression from Eq.~\eqref{eq:baralpha} is shown (blue dashed line), where, for simplicity, the term $\kappa$ is neglected.
As quantitatively described in App.~\ref{app:DOS_approximation}, the agreement of the two curves establishes that, in the extremely weak coupling, $|g|\ll 1$, the instability arises primarily from the presence of modes with total transmission within the GJJ. Their presence makes a Taylor expansion of Eq.~\eqref{eq:baralpha} impossible around $g =0$.

In what follows, we analyze the instability at $\vphi = \pi$ observed in the GJJ case, considering the effects of temperature and the Fermi level. Fig.~\ref{fig:I_pi_vs_mu_T_comparison} shows the supercurrent $I$ at $\vphi = \pi^-$ as a function of the Fermi level $\mu_0$ for three different temperatures: $\kBT  = 0$ (blue solid line), $\kBT / \deltaO = 0.002$ (green solid line), and $\kBT / \deltaO = 0.01$ (red solid line). The other parameters are $\hbar \omega_{\mathrm{r}} = 0.6 \deltaO$ and $g = 0.1$. 
We start by focusing on the case at zero temperature, and we see that the amplitude of the supercurrent is enhanced when the Fermi level $\mu_0$ is tuned close to integer multiples of $\pi \hbar v_{\mathrm{F}} / L$ (vertical dashed lines).
Although the total transmission arising from the Klein tunneling is independent of the Fermi level, the number of values of $k$ that satisfy the stationary-wave condition depends on the Fermi level according to the step function $2 \lfloor |\mu_0|L / (\pi \hbar v_{\mathrm{F}}) \rfloor$. 
Thus, as the Fermi level $\mu_0$ increases, the appearance of additional modes fulfilling the stationary-wave condition leads to an enhancement of the supercurrent~\cite{borzenets2016}.

Considering the expression $I = {\cal N} {\cal P}/(g\phi_0)$, and using that for small $g$ the mean-field ${\cal D}$ is  accurately represented by its value at  $\alpha=0$, ${\cal D}|_{\alpha=0}$ (as shown in Fig.~\ref{fig:L_vs_phi_g_comparison}), one can combine this observation with Eq.~\eqref{eq:MF_alpha-PD}. 
At $\vphi=\pi$, the supercurrent $I$ is therefore proportional to the value of $\bar\alpha$ that minimizes the energy density ${\cal E}(\pi)$. 
Using Eq.~\eqref{eq:baralpha} for $\bar\alpha$, it follows that increasing $\rho_0$ leads to a larger value of $|\bar{\alpha}|$, and thus to an enhanced supercurrent intensity at $\vphi = \pi$.
From the definition $\rho_0 = \sum_{i=-N}^{N}\sqrt{2/|c_{M,i}|}/2\pi$, with $c_{M,i}=\eval{\partial_{k}^2 \tau(k)}_{k_i}$ (see App.~\ref{app:DOS_approximation} for details), where $k_i$ with $i\neq0$ are the momenta which satisfy the stationary-wave condition and  $N=\lfloor |\mu_0|L / (\pi \hbar v_{\mathrm{F}}) \rfloor$,
it therefore follows that the emergence of additional perfectly transparent modes gives additional contributions to $\rho_0$, increasing its value.
However, in Fig.~\ref{fig:I_pi_vs_mu_T_comparison}, one sees a small deviation between the values of $\mu_0$, indicated as $n \pi \hbar v_{\rm F}/L$, and the exact positions of the supercurrent maxima.
Within Eq.~\eqref{eq:baralpha}, this deviation can be traced back to the correction term $\kappa$, which depends on the Fermi level $\mu_0$. App.~\ref{app:DOS_approximation} provides a quantitative analysis accounting for the small offset between the supercurrent peak, as a function of $\mu_0$, and the appearance of further values of $k$ satisfying the stationary-wave condition.

In addition, Fig.~\ref{fig:I_pi_vs_mu_T_comparison} shows that a finite temperature induces a partial suppression of the supercurrent. In particular, at $k_{\rm B}T = 0.01 \Delta_0$ the supercurrent remains appreciable in the vicinity of those values of $\mu_0\gtrsim 15 \hbar v_{\rm F}/L$ and where the zero-temperature supercurrent exhibits a pronounced peak.
As discussed, in a wide and short 2D material-based junction coupled to an LC quantum harmonic oscillator, a finite supercurrent can arise at $\varphi=\pi$ at zero temperature when the coupling constant satisfies $|g|>g_{\rm c}$. In this regime, the coupling between the Andreev system and the photonic mean field $\alpha$ gives rise to new quasiparticle pairs, expressed as linear combinations of ABSs labeled by $k$, with energies $\pm E(k,\varphi)$. At zero temperature, the supercurrent is carried by quasiparticles occupying the $-E(k,\varphi)$ branch.
%
As schematically illustrated in Fig.~\ref{fig:MF_occupancy}, upon increasing the temperature, quasiparticle states with energy $+E(k,\varphi)$ become thermally populated. Since these states carry a supercurrent opposite to that associated with the $-E(k,\varphi)$ branch, their occupation progressively suppresses the net supercurrent. This thermal activation is particularly effective for wavevector components associated with large transmission $\tau(k)$, $\tau(k)\lesssim 1$, because the corresponding energies $E(k,\varphi)$ lie closer to the Fermi level~\cite{pellegrino2022effect}. As noted above, these are also the values of $k$ that make the dominant contribution to the TRB instability, which accounts for the pronounced sensitivity of the instability to the finite temperature.
In the specific GJJ case,  where $g_{\rm c}$ is infinitesimal, Fig.~\ref{fig:I_pi_vs_mu_T_comparison} shows that the supercurrent maintains its robustness against temperature whenever the Fermi level $\mu_0$ assumes values that allow for an extremely large number of ABS pairs associated with perfect or near-perfect transmission. 
This condition corresponds to situations in which the zero-temperature supercurrent at $\vphi=\pi$ exhibits a peak.
At zero temperature, within the mean-field formalism, the energy splitting of the pairs of quasiparticles associated with perfect transmission is $ 2\Delta_0 g|\bar{\alpha}|$,  where $\bar{\alpha}$ is defined in Eq.~\eqref{eq:baralpha}. Consequently, cancellation of the net supercurrent carried by these modes occurs when $k_{\rm B}T \approx \Delta_0 g|\bar{\alpha}|\approx \Delta_0 \exp[-\hbar \omega_{\rm r} /(g^2 n_{\rm v} \rho_0 \Delta_0)]$, as this condition ensures the thermal population quasiparticles with high transmission of both Andreev subbands. This provides an explanation for the resilience of the supercurrent to thermal suppression at the values of the Fermi level where the zero temperature supercurrent is peaked (and, correspondingly, where $|\bar{\alpha}|$ exhibits local maxima at zero temperature).

\begin{figure}[t]
    \centering
    \includegraphics[width=0.475\textwidth]{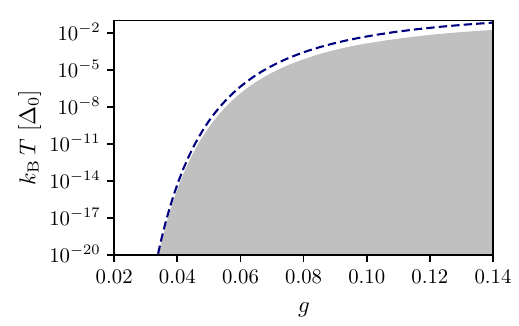}
	\caption[]{%
Phase diagram referring to the spontaneous TRSB instability. 
Gray shaded region represents the values of the coupling constant $g$ and the temperature $T$ (in units of $\Delta_0/k_{\rm B}$) where the global system, composed of GJJ and LC quantum harmonic oscillator, is in TRSB phase, where a finite supercurrent emerges at $\varphi=\pi$, while the white region corresponds to the phase where no supercurrent is present. 
The boundary between the two regions denotes the critical temperature $T_{\mathrm c}$ related to the TRSB instability. The comparison, in logarithmic scale, between the numerical solution and the analytical estimate in Eq.~\eqref{eq:kbT_vs_g_analytical} (blue dashed line) shows that the approximated expression captures the qualitative behavior of the spontaneous instability. 
Here, the other parameters are fixed at $\hbar\omega_{\mathrm r}=0.6\Delta_0$ and $\mu_0 = 10\hbar v_{\mathrm F}/L$. 
        }
	\label{fig:MF_Log_Tc_phase_diagram}
\end{figure}
The critical temperature $T_{\rm c}$ marking the onset of the TRB instability,
consisting in a finite supercurrent at $\vphi=\pi$, can be obtained self-consistently (see App.~\ref{app:Self-consistent problem}). This occurs when the mean-field $\alpha$ assumes a nonzero value, albeit infinitesimal, and contemporary $\mathcal{P}$ also becomes nonzero. 
The result reads
\begin{equation}
    \frac{\hbar\omega_{\mathrm r}}{\deltaO} = \frac{n_{\rm v}g^2}{2} \int_{0}^{1} d\tau\dfrac{\rho(\tau) \tau}{\sqrt{1-\tau}} \Mtanh{\nu_{\rm c} \sqrt{1-\tau}} ~,
	\label{eq:MF_critical_temperature_implicit_rho}
\end{equation}
which implicitly defines the critical temperature in terms of $\nu_{\rm c}=\Delta_0/(2k_{\rm B} T_{\rm c})$.
Focusing on the case of GJJ, isolating the square-root divergence of $\rho(\tau)$ at $\tau=1$, we rewrite the DOS as $\rho(\tau)=\rho_0/\sqrt{1-\tau}+\rho_1(\tau)$. As a result, Eq.~\eqref{eq:MF_critical_temperature_implicit_rho} becomes
\begin{equation}
    \frac{\hbar\omega_{\mathrm r}}{\deltaO} = n_{\rm v}g^2 \rho_0  [\ln(2\nu_{\rm c})+{\cal K}(\nu_{\rm c})]~,
	\label{eq:MF_critical_temperature_GJJ}
\end{equation}
where ${\cal K}={\cal K}_0+{\cal K}_0'+{\cal K}_1$, ${\cal K}_0=\gamma-\ln(\pi/2)-1/2\approx-0.37$, $\gamma$ denotes the Euler–Mascheroni constant,
while both ${\cal K}'_0=\pi^2/(12 \nu_{\rm c}^2)+[\tanh(\nu_{\rm c})-1][\ln(\nu_{\rm c})-1/2]+\int_{\nu_{\rm c}}^{\infty} dz \cosh^{-2}(z)[\ln(z)-z^2/(2 \nu_{\rm c}^2)]$ and ${\cal K}_{1}=(2 \rho_0)^{-1}\int_0^1 d \tau  \Mtanh{\nu_{\rm c} \sqrt{1-\tau}} \rho_1(\tau) \tau/\sqrt{1-\tau}$ are functions of $\nu_{\rm c}$.
To find Eq.~\eqref{eq:MF_critical_temperature_GJJ}, we apply the mathematical manipulations used in deriving Eqs.~\eqref{eq:MF_Tc_DOS_singular} and \eqref{eq:MF_Tc_DOS_singular_2} in App.~\ref{app:DOS_approximation}.
Assuming that the critical temperature $T_{\rm c}$, determined from 
Eq.~\eqref{eq:MF_critical_temperature_GJJ}, satisfies $\nu_{\rm c}\gg 1$, we have 
${\cal K}_0' \to 0$, ${\cal K}_1 \to \kappa_1$, and $\ln(\nu_{\rm c}) \gg 1$. Under these conditions, we obtain the following closed-form approximation for the critical temperature
\begin{equation}\label{eq:kbT_vs_g_analytical}
    k_{\rm B} T_{\rm c} \approx 
    \Delta_0 \exp\left[-\frac{\hbar \omega_{\rm r}}{g^2 n_{\rm v}\rho_0 \Delta_0}\right],
\end{equation}
which confirms the estimate discussed above.
Fig.~\ref{fig:MF_Log_Tc_phase_diagram} shows the phase diagram of the instability that generates 
a finite supercurrent at $\varphi=\pi$ in the GJJ case, corresponding to a time–reversal symmetry breaking 
(TRB) phase. The gray shaded region indicates the TRB phase, while the white area denotes the regime where no supercurrent is expected. The boundary separating the two regions represents the critical 
temperature $T_{\rm c}$ obtained by numerically solving the self-consistent problem, see 
App.~\ref{app:Self-consistent problem} for details. The approximate expression for $T_{\rm c}$, given 
in Eq.~\eqref{eq:kbT_vs_g_analytical}, is shown as the blue dashed line. 
The good agreement in the extremely weak–coupling regime highlights the crucial role played by 
highly transparent modes in driving the instability, while it also emphasizes the fragility of this instability at finite temperatures.

\section{Hybrid excitations}\label{sec:polaritons}
\begin{figure*}[t]
	\begin{overpic}[width=\textwidth]{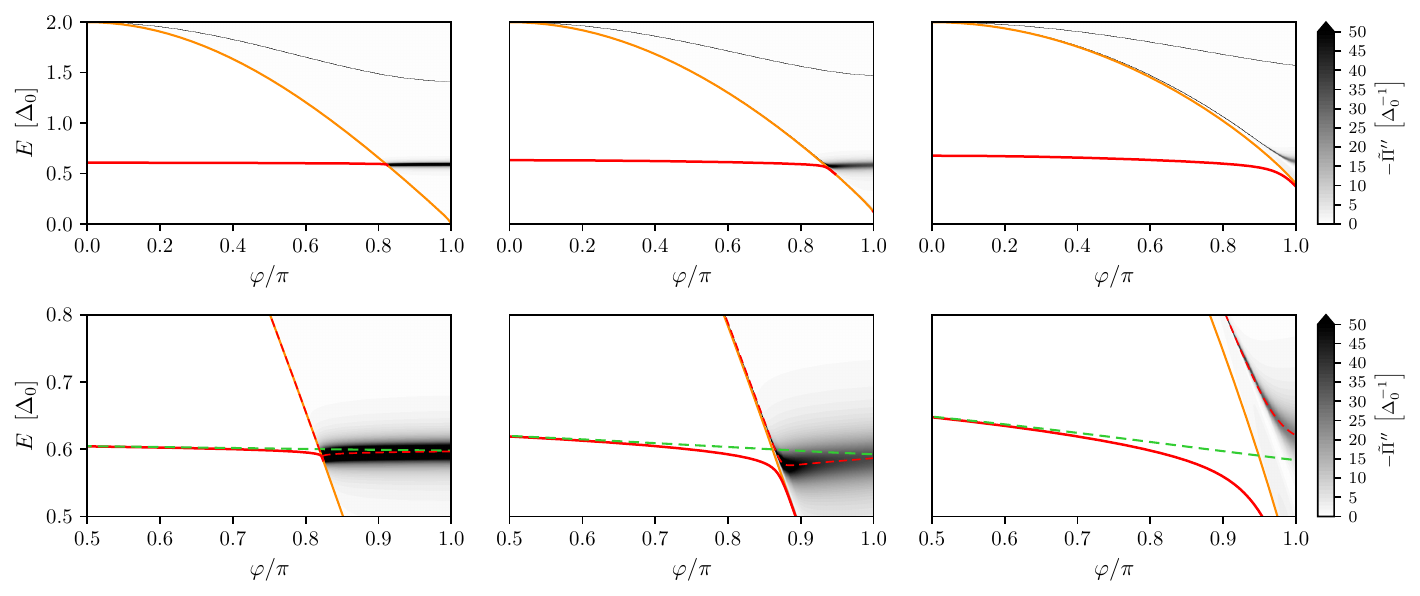}
        \put(0,42){(a)}\put(33.5,42){(b)}\put(63,42){(c)}
        \put(0,21){(d)}\put(33.5,21){(e)}\put(63,21){(f)}
    \end{overpic}
	\caption[]{%
Hybrid excitation spectra of a GJJ coupled with a LC quantum harmonic oscillator as functions of the superconducting phase difference $\varphi$, for three values of the coupling constant, $g=0.1$ (a), $g=0.2$ (b), and $g=0.3$ (c).
The Fermi level is fixed at $\mu_0 = 6.37 \hbar v_{\rm F}/L$ (a), $6.65 \hbar v_{\rm F}/L$ (b), and $7.16 \hbar v_{\rm F}/L$ (c), corresponding for each case to the value of the Fermi level that maximizes the supercurrent at $\varphi=\pi$ for $\mu_0 \gtrsim 2\pi \hbar v_{\rm F}/L$ at zero temperature.
Panels (d)–(f) display corresponding  magnified views around the characteristic LC resonator energy $\hbar\omega_{\rm r}$, limited to the interval $\vphi \in [\pi/2, \pi]$.
The remaining parameters are $\hbar\omega_{\rm r}=0.6\Delta_0$ and $k_{\rm B}T/\Delta_0=0.01$.
In all panels,  the grayscale colormaps display $-\tilde{\Pi}^{\prime\prime}(E/\hbar)$, where $\tilde{\Pi}^{\prime\prime}(E/\hbar)=\Im[\tilde{\Pi}(E/\hbar)]$, and it is nonzero only for $E>2\Delta_{\rm E}(\varphi)$, the orange solid line represents the threshold $2\Delta_{\rm E}(\varphi)$.
Below this threshold, the red solid line shows the low-energy hybrid excitation eigenenergy $\hbar \Omega_n$, obtained as the solution of Eq.~\eqref{eq:MF_linear_ret_response_funct_freq_poles}.
In the zoomed panels, the green dashed lines represent the renormalized bosonic energy $\hbar\lambda\omega_{\rm r}$, whereas the red dashed line denotes the centroid of the most broadened resonance peak, which is a solution of Eq.~\eqref{eq:MF_linear_ret_response_funct_freq_poles} and lies within the support of $\tilde{\Pi}^{\prime\prime}(E/\hbar)$.
        }
	\label{fig:Im_Pi_contourplot_vs_phi_comparison}
\end{figure*}
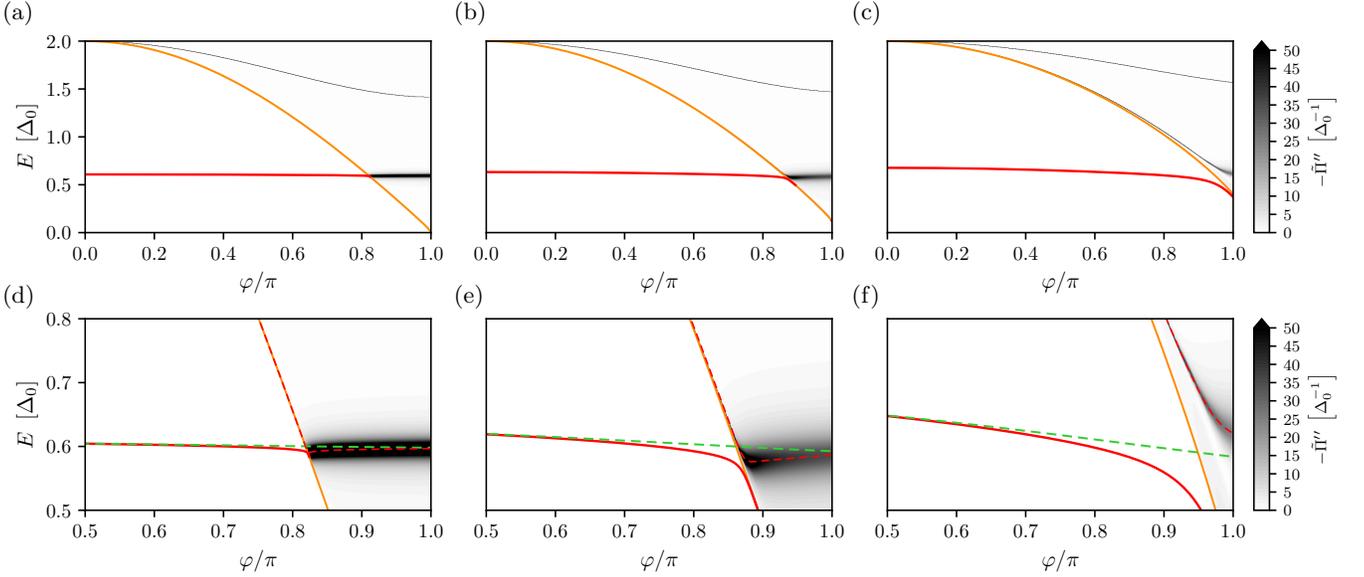

In this Section, we analyze the spectrum of hybridized light-matter excitations within the global system on top of the solution of the mean-field approach~\cite{hopfield1958,ciuti2005quantum}. 
This analysis aims to evaluate the degree of hybridization of the low-energy excitations and the stability of the mean-field solutions. 
To this end, we analyze Gaussian fluctuations around the mean-field solution.
This corresponds to using the Hamiltonian $\hat{H}_W = \hat{H}_{\mathrm{MF}} + \hat{W}$, where $\hat{W}$ accounts for the leading contributions arising from fluctuations.

Our starting point is the mean-field Hamiltonian, discussed in App.~\ref{app:Self-consistent problem}, expressed on the mean-field eigenstate basis as 
\begin{equation}
	\hat{H}_{\rm MF} = \hbar\lambda\omega_{\rm r}\,\Bc\B+   {\cal N}  \sum_{j=\pm}  \int_{-\infty}^\infty \frac{dk}{2\pi} jE(k,\vphi)\PSidag{j,k}\PSi{j,k}~,
	\label{eq:H_MF_fermionic_diagonalized_discrete}
\end{equation}
here, rewritten up to a constant term. 
Above, $\lambda=\left[1+4\mathcal{D}/(\hbar\omega_{\rm r})\right]^{1/2}$, 
$\hat{\psi}_{j, k}=\cos(\theta_k/2) \hat{\gamma}_{j,k}-j\sin(\theta_k/2) \hat{\gamma}_{-j,k}$, and $\B=\cosh(x)\A+\sinh(x)\A^\dagger+\sqrt{\cal{N}}e^x\alpha$ with $x=\ln(\lambda)/2$.
In the mean-field formalism, fluctuations of the operators around their expectation values are typically disregarded. Here, we reintroduce only those corrections that are linear in the bosonic degrees of freedom
\begin{equation}
	\hat{W} =  \frac{\lambda^{-1/2}}{\sqrt{\mathcal{N}}}(\B+\Bc)\hat{V}_{\rm F}~,
	\label{eq:W_linear_MF_fluctuations}
\end{equation} 
where the operator
\begin{equation}
	\hat{V}_{\rm F} = \left(g \df\hat{H}_{\rm A} - {\cal N}\mathcal{P}\right) - 4\alpha \left( \frac{g}{2}^2\dqf\hat{H}_{\rm A} - {\cal N}\mathcal{D}\right)~,
	\label{eq:W_linear_MF_fluctuations_operator}
\end{equation}
is expressed in terms of the mean-field (fermionic) eigenstate basis as
\begin{equation}
		\hat{V}_{\rm F} =  {\cal N} \sum_{j, \, j^{\prime}  } \int^\infty_{-\infty}   \frac{dk}{2\pi} v_{j,j^{\prime}}(k)  \PSidag{j, k}\PSi{j^{\prime}, k}~,
	\label{eq:W_linear_MF_fluctuations_compact}
\end{equation} 
where the explicit expressions for the matrix elements $v_{j,j^{\prime}}(k)$ are reported in App.~\ref{app:Linear_response}. 

We seek to characterize the spectrum of our system by analyzing its response to an external weak perturbation~\cite{haller2022}. Therefore, in what follows, we focus on the response of the dimensionless coordinate of the quantum LC circuit, $\hat{X} = \A + \Ac$, which is encoded in the fluctuations
\begin{equation}
\delta X(t)  = \expval*{\hat{X}(t)} - \expval*{\hat{X}}_{\rm MF}~.
\end{equation}
Within linear-response formalism, we consider a weak external perturbation added to the system Hamiltonian of the form
\begin{equation}
	\hat{V}(t) = f(t) \left(\A+\Ac\right)~,
	\label{eq:External_weak_flux_probe}
\end{equation}
where $f(t)$ is a function that characterizes its time dependence. It is convenient to express $f(t)$ in terms of its frequency components as
\begin{equation}
f(t)=\int_{-\infty}^{\infty} \frac{d \Omega}{2 \pi} e^{-i \Omega t} \tilde{f}(\Omega)~,
\end{equation}
and for simplicity, we assume its DC component to be zero, i.e. $\tilde{f}(0)=0$.

As detailed in App.~\ref{app:Linear_response}, we find that the linear response of this observable, in frequency domain, is given by 
\begin{equation}
	\begin{aligned}
		\delta\tilde{X}(\Omega)&=\tilde{\Pi}(\Omega)\tilde{f}(\Omega)~,
	\end{aligned}
	\label{eq:MF_linear_response_eq_motion_coord_freq}
\end{equation}
where
\begin{equation}
	\tilde{\Pi}(\Omega) = \frac{2\hbar\omega_{\rm r}}{(\hbar\Omega+i0^+)^2  - (\hbar\lambda\omega_{\rm r})^2 - 2\hbar\omega_{\rm r}\tilde{\chi}(\Omega)}~.
	\label{eq:MF_linear_response_ret_response_funct_freq}
\end{equation}
Here, we note that a crucial role is played by the term
 	\begin{equation}
		\begin{aligned}
		\tilde{\chi}(\Omega) &= \int_{-\infty}^{+\infty} \frac{dk}{2\pi}
        \frac{4 n_{\rm v} E(k,\vphi) \abs{ v_{-,+}(k,\vphi)}^2 \Mtanh{\frac{E(k,\vphi)}{2\kBT}}}{(\hbar \Omega+i0^+)^2-4 E(k,\vphi)^2}~,
		\end{aligned}
		\label{eq:MF_linear_response_chi}
	\end{equation}   
which is related to the current–current susceptibility function of the JJ and is, in general, a complex-valued function $\tilde{\chi}(\Omega) = \tilde{\chi}^{\prime}(\Omega) + i \tilde{\chi}^{\prime\prime}(\Omega)$.
We note that Eq.~\eqref{eq:MF_linear_response_chi} contains an implicit temperature dependance via both the mean-field quasiparticle energies $E(k,\vphi)$ and the matrix elements $v_{-,+}(k,\vphi)$.
The imaginary component $\tilde{\chi}^{\prime \prime}(\Omega)$ can be nonzero only when $\Omega>2\Delta_{\rm E}(\vphi)/\hbar$, 
where $\Delta_{\rm E}(\vphi) = \min_{k}\{E(k,\vphi)\}$ represents a minigap~\cite{banszerus_arxiv_2021,vacante2024impurity}. It defines a phase difference $\vphi$ dependent energy range where quasiparticle excitations are forbidden.
In the case of an isolated JJ ($g=0$), at $\varphi = \pi$, the mini-gap is given by $\Delta_0 \sqrt{1 - \tau_{\rm max}}$, where $\tau_{\rm max} = \max_{k}\tau(k)$. 
When there are total transmission propagating channels, the mini-gap closes at $\varphi = \pi$ because $\tau_{\rm max}=1$. 
However, once the coupling to the LC quantum harmonic oscillator is included, the mini-gap remains finite, provided that the temperature satisfies $T < T_{\rm c}$. 
For GJJs, in particular, $T_{\rm c}$ is finite for any value of $g$, as shown in Eq.~\eqref{eq:kbT_vs_g_analytical}.

The poles of the retarded response function $\tilde{\Pi}(\Omega)$ encode information about the spectrum of hybridized excitations~\cite{dmytruk2024}. Therefore, we evaluate the zeros of $\Re[\tilde{\Pi}(\Omega)^{-1}]$, which corresponds to the solutions of the following non-linear equation
\begin{equation}
    (\hbar\Omega_n)^2 = (\hbar\omega_{\rm r})^2 + 2\hbar\omega_{\rm r}\tilde{\chi}^{\prime}(\Omega_n)~.
    \label{eq:MF_linear_ret_response_funct_freq_poles}
\end{equation}
Real solutions $\Omega_n$ satisfying $\hbar \Omega_n < 2\Delta_{\rm E}(\vphi)$ corresponds to the eigenenergy of a hybridized excitation. 
Conversely, when $\hbar \Omega_n$ is real but fulfills $\hbar \Omega_n \ge 2\Delta_E(\varphi)$, it can resonantly create excitations in the mean-field fermionic sector and the response becomes dissipative.
In the thermodynamic limit, ${\cal N}\gg 1$, dissipation is an intrinsic characteristic of the extended matter system, which effectively acts as its own heat bath, since the energy spectrum becomes continuous and energy relaxation occurs over a continuous range of frequencies~\cite{giuliani2005quantum}.
In the latter situation, $\Omega_n$ characterizes a hybrid resonance with finite lifetime.
To estimate the lifetime, we expand the $\tilde{\Pi}(\Omega)^{-1}$ around $\Omega_n$, which gives
\begin{equation}
    \tilde{\Pi}(\Omega)^{-1}\approx \left[\Omega_n/\omega_{\rm r} - \partial_{\Omega}\tilde{\chi}^{\prime}(\Omega_n)\right][\hbar(\Omega-\Omega_n) + i\Gamma_n]~,
	\label{eq:MF_linear_ret_response_funct_freq_imaginary_part_poles}
\end{equation}
where $\Gamma_n$ is the resonance linewidth, explicitly given by 
\begin{equation}
    \Gamma_n =\frac{-(\omega_{\rm r}/\Omega_n)\tilde{\chi}^{\prime\prime}(\Omega_n)}{1 -(\omega_{\rm r}/\Omega_n)\partial_{\Omega}\tilde{\chi}^{\prime}(\Omega_n)}~,
    \label{eq:MF_linear_ret_response_funct_freq_pole_linewidth}
\end{equation}
and is proportional to the inverse of the characteristic excitation lifetime. 
As discussed earlier, here it is clear that this lifetime remains finite only if $\tilde{\chi}^{\prime \prime}(\Omega_n)$ does not vanish, which is the case when $\Omega_n$ satisfies the inequality $\hbar \Omega_n \ge 2\Delta_E(\varphi)$.

Fig.~\ref{fig:Im_Pi_contourplot_vs_phi_comparison} shows the hybrid excitation spectrum as a function of the superconducting phase difference $\vphi$, for three values of the coupling constant, $g = 0.1$ (a), $0.2$ (b), and $0.3$ (c). The panels~(a)–(c) display the full spectra, while correspondingly the panels~(d)–(f) provide zoomed views around the characteristic LC resonator energy~$\hbar\omega_{\rm r}$, restricted to the interval $\varphi \in [\pi/2,\pi]$. All panels use the parameters $\hbar\omega_{\rm r}=0.6\Delta_0$ and finite temperature $k_{\rm B}T/\Delta_0=0.01$.
For each~$g$, the Fermi level $\mu_0$ is chosen so that it maximizes the supercurrent at $\varphi=\pi$ and zero temperature, such that $\mu_0 \gtrsim 2\pi \hbar v_{\rm F}/L$. 
For example, in the case $g=0.1$ analyzed in Fig.~\ref{fig:I_pi_vs_mu_T_comparison}, the specified value of $\mu_0$ coincides with the maximum of the supercurrent (at zero temperature, solid blue curve of Fig.~\ref{fig:I_pi_vs_mu_T_comparison}), located near the gray vertical dashed line marked by $n=2$ in that figure.
In all panels of Fig~\ref{fig:Im_Pi_contourplot_vs_phi_comparison}, the grayscale colormap represents $-\tilde{\Pi}^{\prime\prime}(E/\hbar)=-\Im[\tilde{\Pi}(E/\hbar)]$, explicitly expressed as
\begin{equation}
    \tilde{\Pi}^{\prime\prime}(\Omega)=
    \frac{4(\hbar\omega_{\rm r})^2\tilde{\chi}^{\prime\prime}(\Omega)}
    {\left[(\hbar\Omega)^2 - (\hbar\omega_{\rm r})^2 - 2\hbar\omega_{\rm r}\tilde{\chi}^{\prime}(\Omega)\right]^2  + \left[2\hbar\omega_{\rm r} \tilde{\chi}^{\prime\prime}(\Omega)\right]^2}~,
	\label{eq:MF_linear_ret_response_funct_freq_imaginary_part}
\end{equation}
which is nonzero only for energies above the threshold $2\Delta_{\rm E}(\varphi)$, indicated by the orange solid line. 
The imaginary component $\tilde{\Pi}^{\prime\prime}(\Omega)$ is finite wherever $\tilde{\chi}^{\prime\prime}(\Omega)$ is not vanishing, and the latter is physically associated with the ability of the matter system
to absorb the incident energy $\hbar \Omega$ by creating pairs of electron- and hole-like mean-field quasiparticle excitations.
The orange solid line, corresponding to $2\Delta_{\rm E}(\vphi)$, in panels (b) and (c), remains nonzero at $\vphi=\pi$, since the minigap is finite, contrary to the case shown in panel (a). 
This means that, specifically for $g=0.1$, the chosen finite temperature $T=0.01\Delta_0/k_{\rm B}$ is larger than the critical temperature $T_{\rm c}$ associated with the instability at $\vphi=\pi$ discussed above.
Consistently, the red solid curve in Fig.~\ref{fig:I_pi_vs_mu_T_comparison} shows that, for $T=0.01\Delta_0/k_{\rm B}$  and $g=0.1$, at $\vphi=\pi$, the supercurrent is non-zero for Fermi level $\mu_0 \gtrsim 15 \hbar v_{\rm F}/L$, namely for values significantly larger than those used in Fig.~\ref{fig:Im_Pi_contourplot_vs_phi_comparison}~(a).
Continuing the analysis of Fig.~\ref{fig:Im_Pi_contourplot_vs_phi_comparison}, below the threshold indicated by the orange line, the red curve represents the hybrid excitation eigenenergies $\hbar\Omega_n$. These are poles of the linear response function $\tilde{\Pi}(E/\hbar)$ and are determined by Eq.~\eqref{eq:MF_linear_ret_response_funct_freq_poles}. 
Moreover, in the zoomed-in panels, the green dashed line indicates the renormalized bosonic energy $\hbar\lambda\omega_{\rm r}$, while the red dashed line shows the centroid of the most broadened resonance peak, which is also obtained from Eq.~\eqref{eq:MF_linear_ret_response_funct_freq_poles} and lies within the support of $\tilde{\Pi}^{\prime\prime}(E/\hbar)$.
All solutions $\hbar\Omega_n$ of Eq.~\eqref{eq:MF_linear_ret_response_funct_freq_poles} that satisfy $\hbar\Omega_n \ge 2\Delta_{\rm E}(\varphi)$ correspond to resonances whose linewidth is proportional to $\tilde{\chi}^{\prime\prime}(\Omega_n)$, as shown in Eq.~\eqref{eq:MF_linear_ret_response_funct_freq_pole_linewidth}.
%
%
In Fig.~\ref{fig:Im_Pi_contourplot_vs_phi_comparison}~(a-c), two resonance peaks are visible, each corresponding one-to-one to the square-root singularities of $\tilde{\chi}^{\prime\prime}(\Omega)$ that arise at the two distinct local maxima of $2E(k,\vphi)$.
For completeness, we recall that $\tilde{\chi}^{\prime}(\Omega)$ and $\tilde{\chi}^{\prime\prime}(\Omega)$ are connected by the Kramers-Kr\"onig relations~\cite{GrossoParravicini2013,bruus2004many}. 
Using these relations, one can verify that every square-root divergence in $\tilde{\chi}^{\prime\prime}(\Omega)$ occurring when $\hbar \Omega$ matches a local maximum of $2E(k,\varphi)$ is paired with a corresponding square-root singularity in $\tilde{\chi}^{\prime}(\Omega)$ at the same frequency.
The singular behavior of $\tilde{\chi}^{\prime}(\Omega)$ associated with each local maximum of $2E(k,\varphi)$ ensures that there is a frequency $\Omega_n$ that satisfies Eq.~\eqref{eq:MF_linear_ret_response_funct_freq_poles}.
For any choice of the coupling constant $g$, the highest-energy resonance of $\tilde{\Pi}^{\prime\prime}(\Omega)$ lies at an energy well separated from the bare resonator frequency $\hbar\omega_{\rm r}$, and is essentially found at those values of $\hbar \Omega$ where $\tilde{\chi}^{\prime\prime}(\Omega)$ diverges. 
In this regime, $\tilde{\Pi}^{\prime\prime}(\Omega)$ displays a very narrow linewidth, corresponding to long-lived excitations that are predominantly of matter nature and extremely weakly hybridized with the photonic mode.
Conversely, for each value of $g$, the lower resonance (shown in the zoomed panels as a red dashed line) is shifted upward compared to the value of $\hbar \Omega$ at which $\tilde{\chi}^{\prime\prime}(\Omega)$ diverges within the continuum. 
In addition, this lower resonance exhibits a larger linewidth, which becomes particularly broad when it approaches the bare resonator energy $\hbar\omega_{\rm r}$. 
These features arise from the resulting light-matter hybridization.
Moreover, as the coupling constant $g$ is increased, the level repulsion between the eigenenergy below the threshold (orange solid line) and the lowest-energy resonance just above that threshold becomes increasingly pronounced. 
This behaviour is analogous to the vacuum Rabi splitting reported in planar germanium JJs~\cite{hinderling2024} and indicates a progressive enhancement of light–matter hybridization with stronger coupling.


\section{Conclusions}\label{sec:conclusions}
In this work, we investigated the modifications of the equilibrium properties of a 2D material-based JJ when it is embedded in a superconducting loop that is inductively coupled to a superconducting resonator, focusing on the case of a graphene Josephson junction as an illustrative example. We considered the regime of a wide and short junction and employed a finite-temperature mean-field framework to treat the light-matter coupling.

Our analysis focused on the modifications of the CPR and the hybridized excitation spectrum of the global system. From the CPR, we identified clear signatures of spontaneous TRSB, which we connected to the transmission properties of the 2D electron system in the normal phase. For graphene, we showed how both the Fermi level and finite temperature govern the onset and observability of this instability, and we derived an analytical expression for the corresponding critical temperature. In particular, we found that a sufficiently large number of highly transmissive modes can be essential for the instability to emerge.

The finite supercurrent observed at $\phi=\pi$ was shown to be a direct consequence of spontaneous TRSB occurring in the hybrid system. Here, above the critical coupling, $g>g_{\rm c}$, the global mean-field ground state becomes doubly degenerate, giving rise to two symmetry-related solutions characterized by opposite supercurrents and resonator fluxes, the latter being proportional to $\bar{\alpha}$, namely $(I(\pi)>0,\bar{\alpha}>0)$ and $(I(\pi)<0,\bar{\alpha}<0)$.
This direct correspondence between the electronic and photonic degrees of freedom has no counterpart in an isolated GJJ and represents a distinctive feature of the hybrid light-matter platform considered here.

We further assessed the stability of the mean-field solutions by calculating the hybrid excitation spectrum. For each case, both the hybrid excitation eigenenergies and the damped hybrid resonances were obtained through a linear-response analysis. 
The findings indicate that the degree of light–matter hybridization in low-energy excitations is jointly governed by the light–matter coupling strength, the position of the Fermi level, and the temperature.

\begin{acknowledgments}
	The authors thank G. Anfuso, G.G.N. Angilella, F. Bonasera, G. Chiriacò, L.
	Giannelli, N. Macrì, I. Vacante, for their insightful comments and constructive feedback throughout various stages of this work.
    V.V. and E.P. thank the PNRR MUR project PE0000023-NQSTI.
    E.P. acknowledges support from COST Action CA21144 superqumap. 
    F.M.D.P. acknowledges support from the project PRIN 2022 - 2022XK5CPX (PE3) SoS-QuBa - ``Solid State Quantum Batteries: Characterization and Optimization". 
    G.F. thanks for the support ICSC - Centro Nazionale di Ricerca in High-Performance Computing, Big Data and Quantum Computing under project E63C22001000006, and Universit\`a degli Studi di Catania, 
    project TCMQI PIACERI 2024/2026. 
\end{acknowledgments}

\appendix
\counterwithin{figure}{section}
\section{DERIVATION OF EQ.~\eqref{eq:high_order_term_andreev_continuum}}\label{app:Andreev_basis}
In this appendix, we show a detailed derivation of the expressions for the Andreev current operator, $\hat{I}_{\rm A}$, and the inverse inductance operator, $\hat{\cal L}^{-1}_{\rm A}$, following the procedure described in Ref.~\cite{bretheau2013}. As mentioned in the main text, these operators are related to the first and second derivatives with respect to the superconducting phase difference of the Andreev Hamiltonian $\hat{H}_{\rm A}$, introduced in Eq.~\eqref{eq:high_order_term_andreev_continuum} of the main text.

Without loss of generality, we restrict the discussion to a single $2\times 2$ ABS subspace labeled by $(\zeta,k)$. Since these indices are kept fixed throughout this appendix, they will be omitted from the notation in what follows. 
The full expressions for those operators are subsequently obtained by summing the contributions from all such independent subspaces labeled by $(\zeta,k)$. 
Specifically, we consider the Hamiltonian of a single ABS pair associated with a conduction channel of normal-state transmission probability $\tau$ in a short Josephson junction at an arbitrary superconducting phase difference $\vphip$. 
In the phase-dependent Andreev basis, the Hamiltonian reads
\begin{equation}
	\hat{H}^{\prime}_{\rm A} = \epsilon(\vphip)  \hat{\Upsilon}^\dagger(\vphip) \sigma^z \hat{\Upsilon}(\vphip)~,
	\label{eq:H_Andreev_app}
\end{equation}
where the spinor operator takes the form
\begin{equation}
\hat\Upsilon(\vphi')=
\begin{pmatrix}
\gammaop{+}(\vphip)\\
\gammaop{-}(\vphip)
\end{pmatrix}~,
\end{equation}
$\sigma^z$ denotes the Pauli  $z$-matrix, $\gammaop{j}(\vphip)$ annihilates the ABS with energy above ($j=+$) or below ($j=-$) the Fermi level. 
In this eigenbasis, the Andreev current operator, obtained from the microscopic calculation of Ref.~\cite{pellegrino2022effect}, takes the form
\begin{equation}
    \begin{aligned}
	\hat{I}^{\prime}_{\rm A} &= \frac{1}{\phi_0}\partial_{\vphip}\hat{H}^{\prime}_{\rm A}\\
    &=\frac{\partial_{\vphip}\epsilon(\vphip)}{\phi_0}\hat\Upsilon^\dagger(\vphip)\bigg[\sigma^z  
    - \sqrt{1-\tau}\Mtan{\frac{\vphip}{2}} \sigma^x  \bigg]\hat\Upsilon(\vphip)~,
    \end{aligned}
	\label{eq:Curr_Andreev_app}
\end{equation}
where $\sigma^x$ denotes the Pauli  $x$-matrix.

Now, we introduce the unitary transformation $e^{- i {\Lambda}(\vphi,\vphip)}$, which allows us to express the spinor $\hat{\Upsilon}(\vphip)$, associated with a generic superconducting phase difference $\vphip$, in terms of the spinor corresponding to the fixed phase difference $\vphi$ as
\begin{equation}\label{eq:Ups-Ups}
\hat{\Upsilon}(\vphip)=e^{- i {\Lambda}(\vphi,\vphip)} \hat{\Upsilon}(\vphi)~.
\end{equation}
This ansatz is expected to be valid in the limit $|\vphip-\vphi |\ll 1$. Indeed, the ABSs at the superconducting phase difference $\vphip$ should, in general, be expanded in terms of the complete set of eigenstates at the superconducting phase difference $\vphi$, including both the bound and continuum Andreev states. Neglecting the continuum contribution is justified only when the above condition is satisfied. In all applications of the unitary transformation considered in this Appendix, we explicitly restrict to the regime $|\vphip-\vphi| \ll 1$.

According to this unitary transformation, the Andreev Hamiltonian is expressed as
\begin{equation}
\hat{H}_{\rm A}'=
\hat\Upsilon^\dagger(\vphi)
e^{i {\Lambda}(\vphi,\vphip)}
\epsilon(\vphip)\sigma^z e^{-i {\Lambda}(\vphi,\vphip)}
\hat{\Upsilon}(\vphi)~,
 \label{eq:Andreev_fixed_phase_Hamiltonian_app}
\end{equation}
and similarly, by applying the same approach, the current operator takes the form
\begin{equation}
\begin{aligned}
\hat{I}_{\rm A}'&=\frac{\partial_{\vphip}\epsilon(\vphip)}{\phi_0}\hat\Upsilon^\dagger(\vphi) e^{i {\Lambda}(\vphi,\vphip)}
\Big[\sigma^z - \sqrt{1-\tau}\Mtan{\frac{\vphip}{2}} \sigma^x \Big]
 \\
&\times e^{-i {\Lambda}(\vphi,\vphip)}\hat\Upsilon(\vphi)~.
\label{eq:Curr_Andreev_fixed_phase_app}
\end{aligned}
\end{equation}
By taking the derivative of Eq.~\eqref{eq:Andreev_fixed_phase_Hamiltonian_app} with respect to $\vphip$, one finds
\begin{equation}
\begin{aligned}
	\partial_{\vphip}\hat{H}^{\prime}_{\rm A} &= \hat\Upsilon^\dagger(\vphi) e^{i{\Lambda}(\vphi,\vphip)}\Big\{
    \partial_{\vphip}\epsilon(\vphip)\sigma^z \\
    &+ i\epsilon(\vphip)\comm*{\partial_{\vphip}{\Lambda}(\vphi,\vphip)}{\sigma^z}\Big\}e^{-i{\Lambda}(\vphi,\vphip)}\hat\Upsilon(\vphi)~.
	\label{eq:der_Andreev_fixed_phase_Hamiltonian_app}
\end{aligned}    
\end{equation}
Comparing this expression with Eq.~\eqref{eq:Curr_Andreev_fixed_phase_app}, we have that
\begin{equation}
{\Lambda}(\vphi,\vphip) = {\cal X}(\vphi,\vphip) \sigma^y~,
\end{equation}
where $\sigma^y$ denotes the Pauli  $y$-matrix, and the scalar function \({\cal X}(\vphi,\vphip)\) satisfies
\begin{equation}\label{eq:dX}
\partial_{\vphip}  {\cal X}(\vphi,\vphip) = \frac{\partial_{\vphip}\epsilon(\vphip)}{2\epsilon(\vphip)}\sqrt{1-\tau}\Mtan{\frac{\vphip}{2}}~.
\end{equation}
We further note that ${\cal X}(\vphi,\vphip)$ satisfies ${\cal X}(\vphi,\vphi)=0$. This follows directly from Eq.~\eqref{eq:Ups-Ups}, because in the particular case where $\vphip=\vphi$ the unitary transformation reduces to the identity. Therefore, by integrating Eq.~\eqref{eq:dX} and enforcing the condition ${\cal X}(\vphi,\vphi)=0$, we obtain
\begin{subequations}
\begin{align}
     {\cal X}(\vphi,\vphip) &= \frac{\sqrt{1-\tau} (\vphip - \vphi)}{4} - \frac{1}{2}\left[\varkappa(\vphip) -\varkappa(\vphi)\right]~,\\
     \varkappa(u)&= \arctan{\left(\sqrt{1-\tau}\tan{\left(\frac{u}{2}\right)}\right)}~.
\end{align}
\end{subequations}
We can now evaluate the $n$-th derivative of the Hamiltonian expressed in Eq.~\eqref{eq:Andreev_fixed_phase_Hamiltonian_app} with respect to the superconducting phase difference $\vphi$. To do so, we first compute the $n$-th derivative with respect to $\vphip$, express the outcome in terms of $\epsilon(\vphip)$, ${\cal X}(\vphi,\vphip)$, and their derivatives, and only afterwards take the limit $\vphip\to\vphi$.
Restricting ourselves to the first and second derivatives, by using $\hat{\sigma}^{z/x}(\vphi)=\sum_{j=\pm,j'=\pm}\gammaopdag{j}(\vphi) \sigma^{z/x}_{j j'}\gammaop{j'}(\vphi)$, we finally obtain
\begin{equation}
    \begin{aligned}
    \partial_{\vphi}\hat{H}_{\rm A} &= \lim_{\vphip\to\vphi} \partial_{\vphip}\hat{H}'_{\rm A}\\
    &=\partial_{\vphi}\epsilon(\vphi)\left[\sigmaz(\vphi)- \sqrt{1-\tau}\Mtan{\frac{\vphi}{2}} \sigmax(\vphi)\right]~,\\
   \partial^2_{\vphi}\hat{H}_{\rm A}  &=\lim_{\vphip \to \vphi}\partial^2_{\vphip} \hat{H}^{\prime}_{\rm A} \\
    &= \df\epsilon(\vphi)\times\\
    &\times\left[\frac{\tau +(2-\tau) \Mcos{\vphi}}{2\Msin{\vphi}}\, \sigmaz(\vphi) - \sqrt{1-\tau}\sigmax(\vphi)\right]~,
    \end{aligned}
\end{equation}
which are the expressions given in Eq.~\eqref{eq:high_order_term_andreev_continuum} of the main text.
\section{SELF-CONSISTENT PROBLEM}\label{app:Self-consistent problem}
\begin{figure*}[t]
	\subfloat[]{
		\begin{minipage}[t]{0.48\textwidth}
			\centering
			\includegraphics[width=\textwidth]{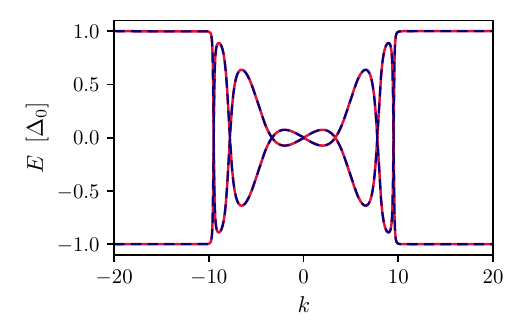}
			\label{fig:MF_spectrum} 
		\end{minipage}
	}\hfill
	\subfloat[]{
		\begin{minipage}[t]{0.48\textwidth}
        \includegraphics[width=\textwidth]{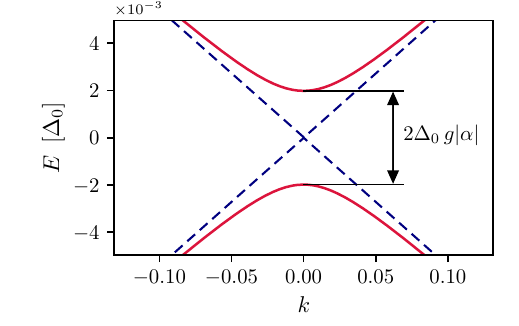}
			\label{fig:MF_spectrum_subgap}
		\end{minipage}	
	}
	\caption[]{%
		(a) Dispersion relations of the mean-field quasiparticles $\pm E(k,\pi)$, as defined in Eq.~\eqref{eq:MF_fermionic_spectrum}, for a short and wide GJJ coupled to an LC quantum harmonic oscillator (red solid line) with coupling constant set at $g=0.1$, compared to the corresponding dispersion relations $\pm \epsilon(k,\pi)$ of the same, but now isolated, GJJ ($g=0$).
        (b) Zoomed-in view of the dispersion relations near $k_0=0$. In this region, an energy splitting with amplitude $2\Delta_{\rm E}=2\deltaO g\abs{\alpha}$ emerges, and an analogous splitting occurs around each value $k_i$ associated with perfect transmission ($\tau(k_i)=1$). In both panels, the remaining parameters are fixed to $\hbar\omega_{\rm r} = 0.6\deltaO$, $\mu_0 = 10 \hvf/L$, and $T=0$.
    }
	\label{fig:MF_spectrum_results}
\end{figure*}

In this Appendix, we derive the explicit expression for the self-consistent system of equations introduced in Eqs.~\eqref{eq:MF_terms} of the main text.
In what follows, we adopt the following notation 
\begin{equation*}
	\expval*{\hat{O}}_{\rm MF} = 
		\dfrac{\Tr( \hat{\rho}_{\rm MF} \hat{O})}{ {\cal Z}_{\rm MF} }~,
\end{equation*}
where $\hat{O}$ is a generic system operator and $\beta=(\kBT)^{-1}$. The mean-field thermal density matrix is given by $\hat{\rho}_{\rm MF}=e^{- \beta \hat{H}_{\rm MF} }$, and the mean-field partition function is represented as ${\cal Z}_{\rm MF}=  \Tr( \hat{\rho}_{\rm MF} ) $. The trace operation $\Tr(\cdot)$ is performed over both the fermionic ABSs sector and the photonic degrees of freedom.
In the zero-temperature case, this thermal average tends to the expectation value in the mean-field factorized ground state $\ket{\Psi_{0}} = \ket{B_0}\ket{F_0}$.
Where $\ket{B_0}$ and $\ket{F_0}$ are, respectively, the ground-state of the Hamiltonians $\hat{H}_{\rm B}$ and $\hat{H}_{\rm F}$, reported in  Eqs.~\eqref{eq:H_MF_terms} of the main text.

We obtain the set of equations by formulating and analyzing the eigenvalue problems associated with the bosonic Hamiltonian $\hat{H}_{\rm B}$ and the fermionic Hamiltonian $\hat{H}_{\rm F}$.
The former can be diagonalized via the Bogoliubov–Valatin transformations
\begin{equation}
	\A= \Mcosh{x}\B - \Msinh{x}\Bc - \sqrt{\mathcal{N}}\alpha~,
	\label{eq:MF_bosonic_BV_transformation}
\end{equation}
where with $\ket{B_0}$ we denotes the vacuum state of $\B$.  During the diagonalization procedure, imposing that the terms proportional to $\B^2$ ($(\Bc)^2$) and $\B$ ($\Bc$) vanish yields
\begin{subequations}
	\begin{align}
     x &= \frac{1}{2}\ln(\lambda)~, \label{eq:x_expression}\\
      \lambda &= \left[1 + \dfrac{4\mathcal{D}}{ \omega_0}\right]^{1/2}~,\label{eq:lambda_expression}\\
		\alpha&=\frac{\mathcal{P}}{\hbar\omega_{\rm r} + \mathcal{D}}~.\label{eq:alpha_expression}
	\end{align}
\label{eq:MF_bosonic_parameters}
\end{subequations}
The resulting diagonal mean-field bosonic Hamiltonian takes the form
\begin{equation}
	\hat{H}_{\rm B} = \hbar\lambda\omega_{\rm r}\left(\Bc\B + \frac{1-2\lambda\alpha^2}{2}\right)~,
	\label{eq:H_MF_bosonic_diagonalized}
\end{equation}
where both the characteristic frequency and the zero-point energy are modified by the coupling. 

Using the diagonal Hamiltonian in Eq.~\eqref{eq:H_MF_bosonic_diagonalized}, the thermal averages the operator $\B$ and $\Bc$ are readily evaluated as
\begin{equation}
	\expval*{\B}_{\rm MF} = \expval*{\Bc}_{\rm MF} = 0~,
	\label{eq:Bosonic_mean_field_termal_avrg}
\end{equation}
from which, in the limit $\mathcal{N}\gg1$, one obtains
\begin{equation}
    \begin{aligned}
        \expval*{\A + \Ac}_{\rm MF} &=-2\sqrt{\mathcal{N}}\alpha~,\\
        \expval*{(\A + \Ac)^2}_{\rm MF} &= 4\mathcal{N}\alpha^2~.
    \end{aligned}
    \label{eq:bosonic_mean_f_exp_values}
\end{equation}
On the other hand, the fermionic Hamiltonian $\hat{H}_{\rm F}$, shown in Eq.~(\ref{eq:H_MF_fermionic}) of the main text, can be conveniently rewritten in the following compact form
\begin{equation}
\hat{H}_{\rm F}= \mathcal{N} \int_{-\infty}^{+\infty} \frac{dk}{2\pi}\, \hat{h}_{k}~,
	 \label{eq:H_MF_fermionic_rewritten}
\end{equation}
where $\hat{h}_{k}=\vd{k}\cdot\hat{\vb*{\sigma}}_{k}$ is a $2\times2$ block Hamiltonian which acts in the subspace spanned by the ABSs with fixed $k$.
In particular, it mimics the structure of the interaction term between a fictitious couple of spin and magnetic field. Here, we introduce $\hat{\vb*{\sigma}}_{k}=(\sigmaz_{k},\sigmax_{k})^{\rm T}$ as a vector of Pauli operators and $\vd{k} =(\cdz{k},\cdx{k})^{\rm T}$ that represents the vector field. The components and the modulus of this field are obtained from Eq.~\eqref{eq:high_order_term_andreev_continuum} and~\eqref{eq:H_MF_fermionic} of the main text
\begin{widetext}
\begin{subequations}
	\begin{align}
		\cdx{k} &=  2g\alpha\df\epsilon(k,\vphi)\sqrt{1-\tau(k)}\bigg(\Mtan{\frac{\vphi}{2}} - g\alpha \bigg)~,\\
		\cdz{k} &= \epsilon(k,\vphi) - 2g\alpha\df\epsilon(k,\vphi)\bigg(1 -  g\alpha\frac{\tau(k) +(2-\tau(k)) \Mcos{\vphi}}{2\Msin{\vphi}}\bigg)~,\\
		E(k,\vphi)&=\abs{\vd{k}}=\sqrt{\cdx{k}^2+\cdz{k}^2}  ~.\label{eq:MF_fermionic_spectrum}
	\end{align}
	\label{eq:MF_d_coefficients}
\end{subequations}
\end{widetext}
Exploiting the fact that the mean-field fermionic Hamiltonian in Eq.~\eqref{eq:H_MF_fermionic_rewritten} can be decomposed into a sum of operators acting on two-dimensional Hilbert subspaces, its diagonalization is simplified, yielding eigenenergies $\pm E(k,\vphi) = \pm \abs{\vd{k}}$.

Accordingly, the mean-field matter Hamiltonian can be written as
\begin{equation}
\hat{H}_{\rm F}= {\cal N}\sum_{j}\int_{-\infty}^{+\infty} \frac{dk}{2\pi}\,jE(k,\vphi)\PSidag{j,k}\PSi{j,k}~,
\label{eq:H_MF_fermionic_diagonalized}
\end{equation}
where $\hat{\psi}^{\dag}_{j,k}$ ($\hat{\psi}_{j,k}$) denotes the fermionic creation (annihilation) operator of quasiparticles labeled with $k$ and energy lying above ($j=+$) or below ($j=-$) the Fermi level. Consequently, the mean-field matter ground state is obtained by filling the vacuum with all quasiparticle states whose energies are below the Fermi level
\begin{equation}
	\ket{F_0} = \prod_{k} \Psidagdw{k}\ket{\rm vac}~.
	\label{eq:Fermionic_mean_field_GS}
\end{equation}
In particular, these operators are derived from the ABSs operators introduced in the main text by applying the following unitary transformation within each $2\times2$ subspace
\begin{equation}
 \mqty(\Psidagup{k} \\  \Psidagdw{k}) = \mqty( \Mcos{\theta_{k}/2} & -\Msin{\theta_{k}/2} \\   \Msin{\theta_{k}/2}& \Mcos{\theta_{k}/2}) \mqty(\gammadagup{k} \\  \gammadagdw{k})~,
 \label{eq:Fermionic_mean_field_quasip}
\end{equation}
where the angle $\theta_k$ is related to the equalities $\Mcos{\theta_{k}}=\cdz{k}/E(k,\vphi)$ and  $\Msin{\theta_{k}}=-\cdx{k}/E(k,\vphi)$.

To evaluate the mean fields introduced in Eqs.~\eqref{eq:MF_terms} of the main text, we exploit the fact that the quasiparticles $\PSidag{j,k}$ obey fermionic statistics. This yields
\begin{equation}
	\expval*{\PSidag{j,k}\PSi{j^{\prime},k}}_{\rm MF} = \krodelta{j,j^{\prime}}n_{\rm F}(jE(k,\vphi))~,
	\label{eq:Fermionic_mean_field_termal_avrg}
\end{equation}
where $\delta_{j,j^\prime}$ denotes the Kronecker delta and $n_{\rm F}(x) = 1/(e^{\beta x}+1)$ is the Fermi–Dirac distribution.
By combining Eqs.~\eqref{eq:MF_terms} of the main text and Eqs.~\eqref{eq:Fermionic_mean_field_quasip} and~\eqref{eq:Fermionic_mean_field_termal_avrg}, after some algebraic manipulations, one gets
\begin{widetext}
\begin{subequations}
	\begin{align}
		\mathcal{P} &=-n_{\rm v}g\int_{-\infty}^{+\infty} \frac{dk}{2\pi} \df\epsilon(k,\vphi) \left[\frac{\cdz{k}}{E(k,\vphi)} - \sqrt{1-\tau(k)}\Mtan{\frac{\vphi}{2}}\frac{\cdx{k}}{E(k,\vphi)}\right]\Mtanh{\frac{E(k,\vphi)}{2k_{\rm B}T}} ~,\label{eq:MF_P_explicit}\\
		\mathcal{D} &= -\frac{n_{\rm v}g^2}{2} \int_{-\infty}^{+\infty} \frac{dk}{2\pi} \df\epsilon(k,\vphi) \left[\frac{\tau(k) +(2-\tau(k)) \Mcos{\vphi}}{2\Msin{\vphi}}\, \frac{\cdz{k}}{E(k,\vphi)}-\sqrt{1-\tau(k)}\,\frac{\cdx{k}}{E(k,\vphi)}\right]\Mtanh{\frac{E(k,\vphi)}{2k_{\rm B}T}} ~.\label{eq:MF_D_explicit}
	\end{align}	
	\label{eq:MF_terms_explicit}
\end{subequations}
\end{widetext}
Here, we introduce the possibility of an additional degree of freedom, which gives rise to a degeneracy factor $n_{\rm v}$. For example, in GJJ the ABS spectrum is doubly degenerate due to the valley degree of freedom, so that $n_{\rm v}=2$.

From Eqs.~\eqref{eq:MF_terms_explicit}, one can straightforwardly obtain the expression for the critical temperature $T_{\rm c}$ given in Eq.~\eqref{eq:MF_critical_temperature_implicit_rho} in the main text.
In fact, for $\vphi=\pi^{-}$ and as the temperature approaches $T_{\rm c}$, time-reversal symmetry requires the supercurrent $I$, and therefore $\alpha$, to tend continuously to $0^{+}$. Hence, imposing the condition $\alpha\approx0^+$ in Eqs.~\eqref{eq:MF_terms_explicit}, evaluated at $\vphi=\pi$, one obtains the following two equations
\begin{subequations}
    \begin{align}
        \frac{\hbar\omega_{\mathrm r}+4\mathcal{D}|_{\alpha=0}}{n_{\rm v}g^2 \deltaO} &=\int_{-\infty}^{+\infty} \frac{dk}{2\pi}\frac{\tau^{2}(k)   \tanh( \nu_{\rm c}\sqrt{1-\tau(k)} ) }{2\sqrt{1-\tau(k)}}~,\\
        -\frac{4\mathcal{D}|_{\alpha=0}}{n_{\rm v}g^2 \deltaO}&= \int_{-\infty}^{+\infty} \frac{dk}{2\pi}
        \frac{\tau(k)\sqrt{1-\tau(k)}  }{2} 
        \\
        &\times\tanh( \nu_{\rm c}\sqrt{1-\tau(k)} ) \nonumber ~,
    \end{align}	
\label{eq:Self_cons_probl_case_phi_pi_critical_temp}
\end{subequations}
where $\nu_{\rm c}=\Delta_0/(2 k_{\rm B} T_{\rm c})$.
By summing these two equations, we obtain
\begin{equation}
\frac{\hbar\omega_{\mathrm r}}{\deltaO} = \frac{n_{\rm v}g^2}{2}\int_{-\infty}^{+\infty} \frac{dk}{2\pi}\,\dfrac{\tau(k)}{\sqrt{1-\tau(k)}} \Mtanh{\frac{\Delta_0\sqrt{1-\tau(k)}}{2\kBTc}} ~,
	\label{eq:MF_critial_temperature_implicit_appendix}
\end{equation}
which coincides with Eq.~\eqref{eq:MF_critical_temperature_implicit_rho} of the main text.

We conclude by examining how the inductive interaction modifies the ABS spectrum, which characterizes a short and wide GJJ, specifically for $\vphi = \pi$, where the time-reversal symmetry breaking phase is fully developed at $\vphi=\pi$. To this end, we set $T = 0$ and choose a finite coupling constant, $g = 0.1$. 
Fig.~\ref{fig:MF_spectrum_results} compares the mean-field quasiparticle spectrum $E(k,\pi)$ (red solid line), which self-consistently depends on the solutions of Eqs.~\eqref{eq:MF_terms_explicit}, and the ABSs spectrum $\epsilon(k,\pi)$ (blue dashed line), reported in Eq.~\eqref{eq:ABSs_energy}. 
For an isolated wide short GJJ, the ABSs spectrum shows a band touching between $\pm \epsilon(k,\pi)$ at those values $k_i$ for which the corresponding transmission equals one, $\tau(k_i)=1$. Conversely , when the GJJ is coupled to the LC quantum harmonic oscillator, these degeneracies are removed due to the inductive coupling between the superconducting loop and the LC circuit. As a representative example, Fig.~\ref{fig:MF_spectrum_results}~(b) shows the scenario close $k_0=0$, here an energy splitting appears with magnitude twice the minigap, $\Delta_{\rm E} = \min_{k}\{E(k)\}=\deltaO g\abs{\alpha}$.

\section{DENSITY OF STATES RESOLVED IN TRANSMISSION PROBABILITY}\label{app:DOS_approximation}
\begin{figure}[t]
\centering
	\includegraphics[width = 0.48\textwidth]{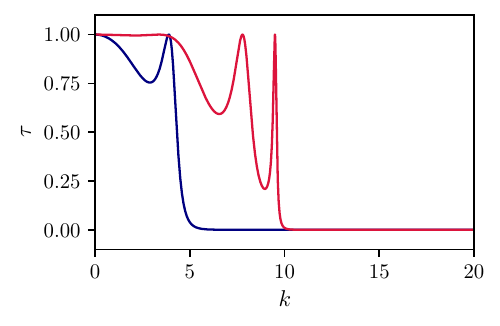}		
	\caption[]{
Transmission probability of graphene in the normal phase, shown for two distinct doping levels, $\mu_0 = 5\hvf/L$ (blue solid curve) and $\mu_0 = 10\hvf/L$ (red solid curve).  Exploiting the even symmetry of the transmission probability, the plot is limited to  $k>0$.
 }
\label{fig:Tau_vs_k}
\end{figure}

\begin{figure*}[t]
	\subfloat[]{
		\begin{minipage}[t]{0.48\textwidth}
			\centering
			\includegraphics[width=\textwidth]{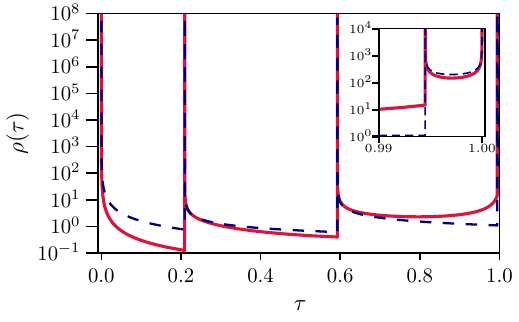}
			\label{fig:rho_comparison_vs_tau} 
		\end{minipage}
	}\hfill
	\subfloat[]{
		\begin{minipage}[t]{0.48\textwidth}
        \includegraphics[width=\textwidth]{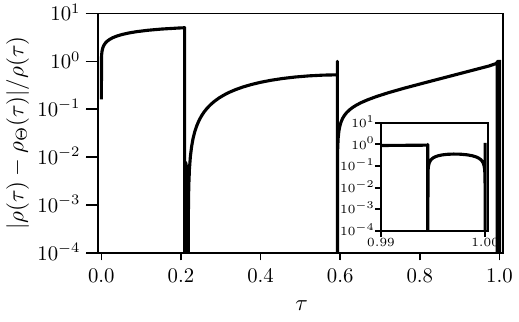}
			\label{fig:rho_comparison_error_vs_tau}
		\end{minipage}	
	}
    \caption[]{%
    (a) DOS resolved in transmission probability $\tau$, for doping level $\mu_0 = 10\hvf/L$. The numerical evaluation of Eq.~\eqref{eq:DOS_piecewice_aprx} is shown as a red solid line, while the approximated closed-form $\rho_\Theta(\tau)$ is represented in a blue dashed line. 
    (b) Relative deviation between the two evaluations of the DOS reported in \ref{fig:rho_comparison_vs_tau} as a function of $\tau$, for $\mu_0 = 10\hvf/L$. 
    In both panels, there is an insert which shows a zoomed view of the near-unity transmission region, $0.99\le\tau\le1$.
    }
	\label{fig:rho_comparison}
\end{figure*}


All results discussed in Sec.~\ref{sec:MF_theory} of the main text are obtained by expressing the self-consistent problem in terms of the density of states (DOS) resolved in the normal-phase transmission probability,
\begin{equation*}
	\rho(\tau) = \int_{-\infty}^{+\infty}\frac{dk}{2\pi}\,\delta(\tau(k)-\tau)~,
\end{equation*}
where $k$ is in units of $1/L$.
Using this representation, a generic quantity $\cal J$, defined as an integral over $k$-space (as $\cal P$ and $\cal D$), can equivalently be written as
\begin{equation}
	{\cal J} = \int_{-\infty}^{+\infty}\frac{dk}{2\pi} J[\tau(k)] = \int_{0}^{1} d\tau\rho(\tau)J(\tau)~.
	\label{eq:DOS_representation}
\end{equation}
This formulation distinguishes between two distinct physical contributions, $\rho(\tau)$ and $J(\tau)$. The former is determined only by the properties of the 2D material of the stripe in the normal phase, whereas the latter is governed by the architecture of the overall platform. In this way, the approach remains general with respect to the specific 2D material used.

Now, we focus on the specific case in which the 2D material is ultra-clean graphene.
Once  $\tau(k)$ is known, the DOS $\rho(\tau)$ can be easily evaluated numerically. 
In addition, here, we provide a reliable approximate closed-form expression for $\rho(\tau)$, obtained by exploiting the properties of transmission probability, $\tau(k)$,  in graphene in the normal phase.
The transmission probability is an even $\tau(-k)=\tau(k)$ and bounded $0\leq\tau(k)\leq1 $ function that exhibits global maxima at $\tau=1$, and several local minima, as shown in Fig.~\ref{fig:Tau_vs_k}. 
By defining $N=\floor{\kf/\pi}$ ($\floor{x}$ is the integer part of $x$ and $\kf = \mu_0L/\hvf$ is the Fermi wavenumber in units of $1/L$), then the number of global maxima is $2N+1$, and the number of non-zero local minima is $2N$, respectively. 

The set of all global maxima, $\{k_i\}$, contains points located at $k_0=0$ (Klein tunneling) and at $k_i = \sgn(i)\sqrt{\kf^2-(i\pi)^2}$ (stationary wave condition) with $i = \pm1,\dots,\pm N$. Moreover, the set of all local minima, $\{k^{*}_i\}$, includes points which are solutions of the transcendental equation $\sqrt{\kf^2 - k^2} = (k/\kf)^2\Mtan{\sqrt{\kf^2 - k^2}}$, which fulfill $k_{i}<k^{*}_i<k_{i+1}$~\cite{pellegrino2022effect}. As illustrated in Fig.~\ref{fig:Tau_vs_k}, once $\mu_0$ surpasses an integer multiple of $\pi \hvf/L$, an additional stationary wave solution becomes allowed, and consequently a new peak emerges in the transmission probability.

In what follows, we derive an analytical approximation for the DOS. 
We start by defining the set $\{\tau_\ell \}\coloneqq\{\tau_0 = 0,\,\tau_1,\dots\,,\tau_{N},\,\tau_{N+1} = 1\}$ that contains all the distinct values of the extrema $N+2$ of the transmission probability sorted in ascending order. 
Fig.~\ref{fig:rho_comparison_vs_tau} shows that these values correspond to points where the DOS presents integrable divergences~\cite{pellegrino2022effect,GrossoParravicini2013}. 
Accordingly, our approach relies on employing a parabolic approximation of the transmission probability in the vicinity of all global maxima and all non-zero local minima, together with an asymptotic approximation near $\tau_0=0$. The latter is given by $\tau(k)\approx 4e^{-2\abs{k}}$, reflecting its exponential decaying behavior for $\abs{k}>\kf$.
Then, we exploit the following useful mathematical relations
\begin{widetext}
\begin{subequations}
	\begin{align}
	\int_{-\infty}^{+\infty}dk\,\delta\left(4e^{-2\abs{k}}-\tau\right) &= \frac{1}{\tau}\quad \tau_{0}<\tau<\tau_{1}~,\\
	\int_{-\infty}^{+\infty}dk\,\delta\left(\tau_\ell+ \frac{c_\ell}{2}(k-k_\ell)^2-\tau\right) &= 
	\begin{cases}
		\sqrt{\dfrac{2^{\mathstrut}}{\abs{c_\ell}}}\dfrac{1}{\sqrt{\tau-\tau_\ell}}\qquad &k_\ell\in\{k^*_i\},\,\tau_\ell<\tau<\tau_{\ell+1}~,\\\\
		\sqrt{\dfrac{2^{\mathstrut}}{\abs{c_\ell}}}\dfrac{1}{\sqrt{1-\tau}} \qquad &k_\ell\in\{k_i\},\, \tau_N<\tau<1~,	
	\end{cases}
	\end{align}
\end{subequations}
\end{widetext}
where $\tau_\ell=\tau(k_\ell)$, and $c_\ell=\eval{\partial^2_{k}\tau(k)}_{k=k_\ell}^{}$ denotes the concavity of the transmission probability evaluated at $k_\ell$.  
The considerations outlined above can be cast in mathematical form by approximating the DOS as
\begin{equation}
	\begin{aligned}
		\rho_{\Theta}(\tau) &= \left( \frac{\rho_0}{\sqrt{1-\tau}}+ \frac{\rho_{m,N}}{\sqrt{\tau - \tau_{N}}}\right)\Theta(\tau-\tau_{N})\Theta(1-\tau)\,+\\
		&+\sum_{\ell=1}^{N-1}\frac{\rho_{m,\ell}}{\sqrt{\tau - \tau_\ell}}\Theta(\tau-\tau_\ell)\Theta(\tau_{\ell+1}-\tau)\,+\\
		&+\frac{1}{2\pi\tau}\Theta(\tau)\Theta(\tau_1-\tau)~,
	\end{aligned}
\label{eq:DOS_piecewice_aprx}
\end{equation}
where $\Theta(x)$ is the Heaviside step function, $\rho_0 = \sum_{\ell}\sqrt{2/\abs{c_{M,\ell}}}/2\pi$ and $\rho_{m,\ell} = \sqrt{2/\abs{c_{m,\ell}}}/\pi$ are weight factors, in terms of $c_\ell$ evaluated respectively for $k_\ell\in\{k_i\}$ and $k_\ell\in\{k^*_i\}$, which depend on the graphene Fermi level, $\mu_0$. 
Fig.~\ref{fig:rho_comparison_vs_tau} compares DOS $\rho(\tau)$, as a function of $\tau$, obtained with a direct numerical evaluation (solid red solid curve) and the corresponding approximate closed-form expression $\rho_\Theta(\tau)$, given by Eq.~\eqref{eq:DOS_piecewice_aprx}, (blue dashed curve), considering the generic value of the Fermi level $\mu_0=10 \hbar v_{\rm F}/L$.

The analysis of the relative deviation $|\rho(\tau)-\rho_\Theta(\tau)|/\rho(\tau)$, reported in Fig.~\ref{fig:rho_comparison_error_vs_tau}, reveals an excellent agreement in the vicinity of the divergences, which allows for a direct and quantitative evaluation of the corrections to the results presented in Eq.~\eqref{eq:baralpha} and Eq.~\eqref{eq:kbT_vs_g_analytical} of the main text.
As discussed in the self-consistent analysis in the main text, these divergences provide the dominant contributions to the modifications of both the CPR and the hybridized excitation spectrum of the system. 
However, the discrepancy between the approximate closed-form expression $\rho_\Theta(\tau)$ and the exact $\rho(\tau)$ remains non-negligible across the full range of transmission values, including the highly transparent region ($0.99 \le \tau \le 1$), as illustrated in the inset of Fig.~\ref{fig:rho_comparison_error_vs_tau}.
\begin{figure*}[t]
	\subfloat[]{
        \begin{minipage}[t]{0.48\textwidth}
			\centering
			\includegraphics[height=0.225\textheight]{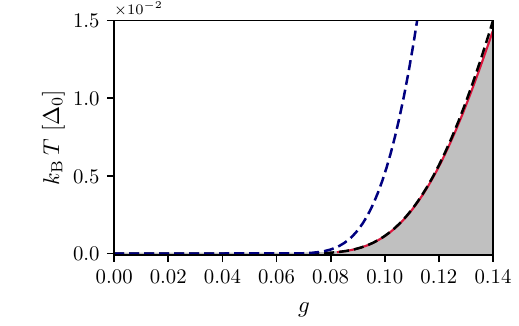}
			\label{fig:MF_Tc_phase_diagram} 
        \end{minipage}
	}\hfill
	\subfloat[]{
        \begin{minipage}[t]{0.48\textwidth}
			\centering
			\includegraphics[height=0.225\textheight]{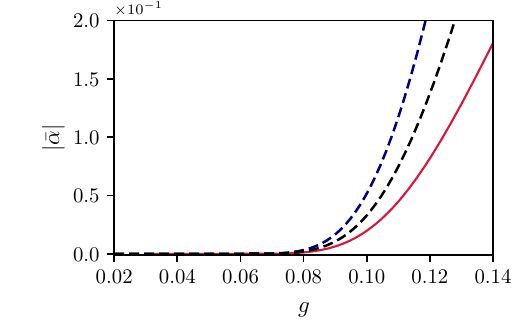}
			\label{fig:alpha_pi_vs_g_comparison_numerical_analytical}
        \end{minipage}
	}
	\caption[]{%
		(a) Phase diagram referring to the TRB instability. Here, in a linear scale, we compare the numerical result for critical temperature $T_{\rm c}$ obtained from Eq.~\eqref{eq:MF_critial_temperature_implicit_appendix} (red solid line) with the two approximations provided by Eq.~\eqref{eq:kbT_vs_g_analytical} of the main text (blue dashed line) and Eq.~\eqref{eq:kbT_vs_g_analytical_appendix} (black dashed line), respectively. 
        (b) Modulus of the photonic mean-field $\bar{\alpha}$ that minimizes the mean-field energy-density functional as a function of the coupling constant at zero temperature. In a linear scale, the full numerical solution of the self-consistent equations (red solid line) is compared with the expression in Eq.~\eqref{eq:baralpha} of the main text evaluated at $\kappa=0$ (blue dashed line) and with Eq.~\eqref{eq:alpha_vs_g_analytical_appendix} (black dashed line). In both panels, we fix $\vphi=\pi$, $\hbar\omega_{\rm r} = 0.6\deltaO$, and $\mu_0 = 10\hvf/L$.
        }
	\label{fig:MF_main_results}
\end{figure*}
We first analyze the critical temperature $T_{\rm c}$ associated with the spontaneous time-reversal symmetry breaking instability discussed in the main text.
Starting from Eq.~\eqref{eq:MF_critical_temperature_implicit_rho} of the main text and employing $\rho_{\Theta}(\tau)$, we obtain the following result
\begin{equation}
	\begin{aligned}
	 \frac{2}{n_{\rm v}g^2}\frac{\hbar\omega_{\mathrm r}}{\deltaO} &=  \rho_0(\mu_0)\int_{\tau_N}^{1} d\tau \dfrac{\tau}{1-\tau}\Mtanh{\nu_{\rm c} \sqrt{1-\tau}}+\,\\
	 &+\sum_{\ell=1}^{N}\rho_{m,\ell}(\mu_0) \int_{\tau_\ell}^{\tau_{\ell+1}}d\tau\frac{\tau}{\sqrt{1 - \tau}\sqrt{\tau - \tau_\ell}}+\,\\
	 &+\frac{1}{2\pi}\int_{0}^{\tau_{1}}d\tau\frac{1}{\sqrt{1 - \tau}}~,
	\end{aligned}
	\label{eq:MF_Tc_DOS_piecewice_aprx_1}
\end{equation}
where $\nu_{\rm c}=\deltaO/(2\kBTc)$.
Since the critical temperature lies deep in the low-temperature regime, $\kBTc \ll \deltaO$, we retain the explicit temperature dependence through the factor $\tanh{\nu_{\rm c}\sqrt{1-\tau}}$ only in the leading term of the DOS approximation. In the remaining terms, $\tanh{(\nu_{\rm c}\sqrt{1-\tau})}$ can be accurately replaced by $1$, because $\nu_{\rm c}\gg 1$ and $0\le \tau <1$.
Consequently, these integrals can be evaluated in analytical closed forms. In particular, the integral reported in the second line of Eq.~\eqref{eq:MF_Tc_DOS_piecewice_aprx_1} reads as follows
\begin{equation}
	\begin{aligned}
    \int_{\tau_\ell}^{\tau_{\ell+1}}d\tau\frac{\tau}{\sqrt{1 - \tau}\sqrt{\tau - \tau_\ell}}&=(1+\tau_\ell)\arcsin({\sqrt{\frac{\tau_{\ell+1}-\tau_\ell}{1-\tau_\ell}}}) \\
    &- 
        \sqrt{(\tau_{\ell+1}-\tau_\ell)(1-\tau_{\ell+1})}~,
    \end{aligned}
    \label{eq:integral_tau_i_tau_i+1}
\end{equation}
and the integral reported in the third line results in 
\begin{equation}
\int_{0}^{\tau_{1}}d\tau\frac{1}{\sqrt{1 - \tau}} = 2 \left[1 - \sqrt{1-\tau_1}\,\right]~.
\label{eq:integral_tau_0_tau_1}
\end{equation}

Then, we focus on the first integral on the right-hand side of Eq.~\eqref{eq:MF_Tc_DOS_piecewice_aprx_1}, and we express it in the form
\begin{equation}
\begin{aligned}
    &\int_{\tau_N}^{1} d\tau \dfrac{\tau}{1-\tau}\Mtanh{\nu_{\rm c}\sqrt{1-\tau}}\\
	&=2\bigg[\int_{0}^{z_N}dz\dfrac{\Mtanh{z}}{z} 
    -  \left(\frac{1}{\nu_{\rm c}}\right)^{2}\int_{0}^{z_N}dz\left(z\Mtanh{z}\right)\bigg]~,
    \label{eq:MF_Tc_DOS_piecewice_aprx_singular}
    \end{aligned}
\end{equation}
where we adopt the substitution $z = \nu_{\rm c}\sqrt{1-\tau}$, which implies $z_N=\nu_{\rm c}\sqrt{1-\tau_{N}}$. 
As a first step~\cite{bruus2004many}, we integrate by parts the first term on the right-hand side of Eq.~\eqref{eq:MF_Tc_DOS_piecewice_aprx_singular}
\begin{equation}\label{eq:MF_Tc_DOS_singular}
	\begin{aligned}
	&\int_{0}^{z_N}dz\dfrac{\Mtanh{z}}{z}=\\
    &=\ln{\left(z_N\right)}\Mtanh{z_N} -\int_{0}^{\infty}dz \frac{\ln{(z)}}{\cosh^{2}{(z)}} +  \int_{z_N}^{\infty}dz \frac{\ln{(z)}}{\cosh^{2}{(z)}}\\
    &=\ln{\left(z_N\right)}\Mtanh{z_N} - \ln{\frac{\pi}{4}} +\gamma +  \int_{z_N}^{\infty}dz \frac{\ln{(z)}}{\cosh^{2}{(z)}}~,
    \end{aligned}
\end{equation}
where $\gamma\approx0.577$ is the Euler-Mascheroni constant. Similarly, we address the second term on the right-hand side of Eq.~\eqref{eq:MF_Tc_DOS_piecewice_aprx_singular}, and we obtain
\begin{equation}\label{eq:MF_Tc_DOS_singular_2}
   \begin{aligned}
   &\left(\frac{1}{\nu_{\rm c}}\right)^{2}\int_{0}^{z_N}dz\left(z\Mtanh{z}\right)=\\
   %
   %
   &= \frac{z_N^2\Mtanh{z_N}}{2\nu^2_{\rm c}}- \frac{\pi}{12\nu^2_{\rm c}} + \frac{1}{2\nu^2_{\rm c}}\int_{z_N}^{\infty}dz \frac{z^2}{\cosh^{2}{(z)}} , \\
   \end{aligned}
\end{equation} 
By combining Eq.~\eqref{eq:MF_Tc_DOS_piecewice_aprx_singular}, Eq.~\eqref{eq:MF_Tc_DOS_singular}, and Eq.~\eqref{eq:MF_Tc_DOS_singular_2} into Eq.~\eqref{eq:MF_Tc_DOS_piecewice_aprx_1}, we obtain
\begin{equation}
    \begin{aligned}
    \frac{1}{g^2}\frac{\hbar\omega_{\mathrm r}}{n_{\rm v}\rho_0(\mu_0)\deltaO} &=   \ln\left(2\nu_{\rm c}\right) + \eta(\mu_0) + \frac{\pi}{12\nu^2_{\rm c}} +\\
    &+ [\tanh(z_N)-1][\ln(z_N)-z^2_N/(2 \nu_{\rm c}^2)]\\
    &+\int_{z_N}^{\infty} dz \frac{\ln(z)-z^2/(2 \nu_{\rm c}^2)}{\cosh^{2}(z)}~.
    \end{aligned}
    \label{eq:MF_Tc_DOS_piecewice_aprx_2}
\end{equation}
Here, we introduce the quantity $\eta(\mu_0)$, where we include the contributions to the critical temperature $T_{\rm c}$
that arise from the local minima of the transmission probabilities in $\rho_\Theta(\tau)$, specifically the terms appearing in the second and third lines of Eq.~\eqref{eq:MF_Tc_DOS_piecewice_aprx_1}.
Since our focus is on the regime $\nu_{\rm c}\gg 1$, we can take the limit $z_N\to\infty$, implying that only the first two terms on the right-hand side of Eq.~\eqref{eq:MF_Tc_DOS_piecewice_aprx_2} need to be retained.
As a result, by solving for $T_{\rm c}$, one obtains
\begin{equation}
	\frac{\kBT_{\mathrm c}}{\deltaO}\approx \exp{-\left[\frac{1}{g^{2}}\frac{\hbar\omega_{\mathrm r}}{n_{\rm v}\rho_0(\mu_0)\deltaO} - \eta(\mu_0)\right]}~.
	\label{eq:kbT_vs_g_analytical_appendix}
\end{equation}

Fig.~\ref{fig:MF_Tc_phase_diagram} shows the phase diagram of the instability that produces a finite supercurrent at $\varphi=\pi$ in a GJJ, indicating a TRB phase. 
The gray area corresponds to the TRB region, while the white region denotes parameters values for which no supercurrent is expected. 
These results are also shown in Fig.~\ref{fig:MF_Log_Tc_phase_diagram} of the main text, where they are presented on a log-scale, in order to highlight the extremely weak-coupling regime, $g\lll1$.
The red solid line represents the boundary that determines the critical temperature $T_{\rm c}$, as obtained from the fully self-consistent numerical solution, while the blue and black dashed lines correspond to the approximate expressions for $T_{\rm c}$ given by Eqs.~\eqref{eq:kbT_vs_g_analytical} and \eqref{eq:kbT_vs_g_analytical_appendix}, respectively.
This comparison indicates that the inaccuracy of the simplified expression in Eq.~\eqref{eq:kbT_vs_g_analytical} primarily arises from the influence of the local minima of the transmission probability, which are entirely disregarded there, yet contribute non-negligibly to the critical temperature for $g \gtrsim 0.05$.

At zero temperature and in the weak-coupling regime ($g\abs{\alpha}\ll1$), by employing the approximate expression of the DOS, $\rho_{\Theta}(\tau)$, we can elucidate how the local minima of the transmission probability affect the photonic mean-field $\bar{\alpha}$, which is determined by minimizing the mean-field energy-density functional given in Eq.~\eqref{eq:baralpha} of the main text.
The starting point is the energy
density functional, defined in Eq.~\eqref{eq:calE} of the main-text, evaluated at $\vphi=\pi$. 
Employing Eq.~\eqref{eq:Ek_pi} of the main text and rewriting the energy density functional in terms of $\rho(\tau)$, as given in Eq.~\eqref{eq:DOS_representation}, 
we obtain
\begin{equation} 
    \begin{aligned}
    \frac{{\cal E}(\pi)}{\deltaO} 
    &=\frac{\hbar\omega_{\rm r}}{\deltaO}\alpha^2 - n_{\rm v}\int_{0}^{1}d\tau \rho(\tau)E(\tau,\pi)\\
    &\approx \frac{\hbar\omega_{\rm r}}{\deltaO}\alpha^2 - n_{\rm v}\int_{0}^{1}d\tau \rho(\tau)\sqrt{1-(1-(g\alpha)^2)\tau}~.
    \end{aligned}
	\label{eq:MF_alpha_implicit_phi_pi}
\end{equation}
Analogously to what was done for the critical temperature, see Eq.~\eqref{eq:MF_Tc_DOS_piecewice_aprx_1}, we replace $\rho(\tau)$ by $\rho_{\Theta}(\tau)$ in the above integral. 
This substitution gives 
\begin{equation}
\begin{aligned}
&n_{\rm v}\int_{0}^{1}d\tau \rho_\Theta(\tau)\sqrt{1-(1-(g\alpha)^2)\tau}\\
&= n_{\rm v}\rho_0(\mu_0)\int_{\tau_N}^{1} d\tau \frac{\sqrt{1-(1-(g\alpha)^2)\tau}}{\sqrt{1-\tau}}\,+\\
	 &+n_{\rm v}\left[\sum_{\ell=1}^{N}\rho_{m,\ell}(\mu_0)\int_{\tau_{\ell}}^{\tau_{\ell+1}}d\tau\frac{\sqrt{1-(1-(g\alpha)^2)\tau}}{\sqrt{\tau - \tau_\ell}}\right.+\,\\
	 &+\left.\frac{1}{2\pi}\int_{0}^{\tau_{1}}d\tau\frac{\sqrt{1-(1-(g\alpha)^2)\tau}}{\sqrt{1 - \tau}}\right]~.
	\end{aligned}
	\label{eq:MF_energy_functional_DOS_piecewice_aprx_1}
\end{equation}
We begin by examining the contribution to $\rho_{\Theta}(\tau)$ which includes the total transmission $\tau=1$, expressed as $\rho_0(\mu_0)/\sqrt{1-\tau}$.
We have
\begin{equation}
	\begin{aligned}
		&\int_{\tau_{N}}^{1}d\tau\dfrac{\sqrt{1-(1-(g\alpha)^2)\tau}}{\sqrt{1-\tau}}\\
		&=\frac{(g\alpha)^2}{\sqrt{1-(g\alpha)^2}}\left[\ln(x_N + \sqrt{x_N^2+1} ) + x_N\sqrt{x_N^2+1}\right]\\
        %
		\label{eq:MF_En_functional_case_phi_pi_rho_zero_exact}
	\end{aligned} 
\end{equation}
where we introduced the change of variables $x = \sqrt{1-(g\alpha)^2}\sqrt{(1-\tau)/(g\alpha)^2}$, which implies $x_N = \sqrt{1-(g\alpha)^2}\sqrt{(1-\tau_N)/(g\alpha)^2}$.
For $g\abs{\alpha} \ll 1$, we retain only the leading-order terms
\begin{equation}
	\begin{aligned}
		&\int_{\tau_{N}}^{1}d\tau\dfrac{\sqrt{1-(1-(g\alpha)^2)\tau}}{\sqrt{1-\tau}} \approx (1-\tau_N)+\\
        %
        %
		& + \left[\frac{\tau_N }{2}+ \ln\left(\frac{2\sqrt{1-\tau_N}}{g\alpha}\right)\right](g\alpha)^2~.
		\label{eq:MF_En_functional_case_phi_pi_rho_zero}
	\end{aligned} 
\end{equation}
We next examine the integrals appearing in the second and third lines on the right-hand side of Eq.~\eqref{eq:MF_energy_functional_DOS_piecewice_aprx_1}.
For each integral, we again retain only the leading contribution in the limit $g |\alpha| \ll 1$.
Exploiting the smoothness of the integrands with respect to $\alpha$ in a neighborhood of $\alpha = 0$, it is sufficient to expand them up to order $(g\alpha)^2$, which yields
\begin{equation}
\begin{aligned}
&n_{\rm v}\int_{0}^{1}d\tau \rho_\Theta(\tau)\sqrt{1-(1-(g\alpha)^2)\tau}\\
&\approx -\varepsilon_0 + n_{\rm v}(g\alpha)^2\left\{\rho_0(\mu_0)\left[\frac{\tau_N }{2}+ \ln\left(\frac{2\sqrt{1-\tau_N}}{g\alpha}\right)\right]\right.+\\
	 &+\sum_{\ell=1}^{N}\frac{\rho_{m,\ell}(\mu_0)}{2}\int_{\tau_\ell}^{\tau_{\ell+1}}d\tau\frac{\tau}{\sqrt{1 - \tau}\sqrt{\tau - \tau_\ell}}+\,\\
	 &+\left.\frac{1}{4\pi}\int_{0}^{\tau_{1}}d\tau\frac{1}{\sqrt{1 - \tau}}\right\}~,
	\end{aligned}
	\label{eq:MF_energy_functional_DOS_piecewice_aprx_2}
\end{equation}
where $\varepsilon_0=- n_{\rm v}\int_{0}^1 d \tau \rho_\Theta(\tau)\sqrt{1-\tau}$.
The remaining definite integrals appearing in Eq.~\eqref{eq:MF_energy_functional_DOS_piecewice_aprx_2} are identical to those defined in Eq.~\eqref{eq:integral_tau_i_tau_i+1} and Eq.~\eqref{eq:integral_tau_0_tau_1}.
Upon substituting all these results into Eq.~\eqref{eq:MF_alpha_implicit_phi_pi}, one finds
\begin{equation}
    	\begin{aligned}
		\frac{{\cal E}(\pi)}{\deltaO}&\approx \varepsilon_0 + \left\{\frac{\hbar\omega_{\rm r}}{\deltaO}+ n_{\rm v}g^2\left[\rho_0(\mu_0)\ln\left(g\alpha\right)-\eta_0(\mu_0)\right]\right\}\alpha^2~,
	\end{aligned}
	\label{eq:MF_En_functional_case_phi_pi_rho_aprx_improved_end}
\end{equation}
where the contributions that arise from the local minima of the transmission probabilities in $\rho_\Theta(\tau)$ are included in the term $\eta_0(\mu_0)$. 
Finally, we solve for the minimum of  Eq.~\eqref{eq:MF_En_functional_case_phi_pi_rho_aprx_improved_end} and obtain
\begin{equation}
   \abs{\bar{\alpha}} = \frac{1}{g}\exp\left\{-\left[\frac{\hbar\omega_{\rm r}}{n_{\rm v}\rho_0(\mu_0)\deltaO g^2} - \eta_1(\mu_0)\right] \right\}~,
	\label{eq:alpha_vs_g_analytical_appendix}
\end{equation}
where and $\eta_1(\mu_0)= (\rho_0(\mu_0))^{-1}\eta_0(\mu_0)-1/2$. 

Fig.~\ref{fig:alpha_pi_vs_g_comparison_numerical_analytical} shows the modulus of the photonic mean field $\bar{\alpha}$ as a function of $g$. The black dashed line is obtained from Eq.~\eqref{eq:alpha_vs_g_analytical_appendix}, the blue dashed line from Eq.~\eqref{eq:baralpha} of the main text evaluated at $\kappa=0$, and the red solid line from the fully self-consistent numerical solution, all computed at zero temperature.
In contrast to Fig.~\ref{fig:alpha_VS_g} of the main-text, the data here are displayed on a linear scale.

Analogously for the critical temperature, the present analysis shows that, as the coupling constant $g$ increases, the contribution of the local minima in the transmission probability becomes increasingly significant for the onset of the instability. 
In contrast to the behavior of the critical temperature shown in Fig.~\ref{fig:MF_Tc_phase_diagram}, Fig.~\ref{fig:alpha_pi_vs_g_comparison_numerical_analytical} demonstrates that the approximation yielding Eq.~\eqref{eq:alpha_vs_g_analytical_appendix} (black dashed line) does not coincide with the numerical data (red solid line).
The origin of this discrepancy can be traced back to the non-regular behavior of the integrand in Eq.~\eqref{eq:MF_alpha_implicit_phi_pi} in the vicinity of $\tau=1$ and $\alpha=0$. In the vicinity of $\alpha = 0$, Eq.~\eqref{eq:MF_alpha_implicit_phi_pi} reduces to the form given in Eq.~\eqref{eq:energy_exp} of the main text. Consequently, due to the square-root singularity of $\rho(\tau)$ in $\tau = 1$, here the integrand in this expression develops a simple pole.
In particular, obtaining a better result requires an accurate treatment of $\rho(\tau) $ in the region $\tau \in [1-(g\bar{\alpha})^2,1]$ where $\bar{\alpha}$ is the solution of the self-consistent problem.
As the magnitude of $g$ increases, this region broadens from the lower side.

On the other hand, for extremely small values of $g\lll 1$, it is sufficient to correctly capture the singular behaviour at $\tau=1$, namely, the square divergence, as we have done using $\rho_\Theta$. However, as $g$ becomes larger, this approximation ceases to provide an equally accurate description.
As shown in the inset of Fig.~\ref{fig:rho_comparison_error_vs_tau}, although $\rho_\Theta(\tau)$ has essentially the same weight on the square root divergence at $\tau=1$ of $\rho(\tau)$  (numerically calculated), it presents a relative deviation with respect to $\rho(\tau)$ also in the highly transparent region. This explains why the discrepancy observed in Fig.~\ref{fig:alpha_pi_vs_g_comparison_numerical_analytical} increases as the intensity of $g$ becomes larger.

This issue does not arise in the determination of the critical temperature, because the presence of the hyperbolic tangent, $\tanh(\nu_{\rm c}\sqrt{1-\tau})$, forces the integrand in Eq.~\eqref{eq:MF_critical_temperature_implicit_rho} to vanish at $\tau=1$, thereby regularizing it in the vicinity of the total transmission. 
Moreover, as $g$ increases, the critical temperature also increases, so $\nu_{\rm c}$ decreases, enlarging the region over which $\tanh (\nu_{\rm c}\sqrt{1-\tau})$ effectively regularizes the integrand. As a result, the critical temperature problem is less sensitive to discrepancies between the approximate density of states $\rho_\Theta(\tau)$ and the exact (numerical) one $\rho(\tau)$.

\begin{figure*}[t]
	\subfloat[]{
		\begin{minipage}[t]{0.47\textwidth}
			\centering
			\includegraphics[width=\textwidth]{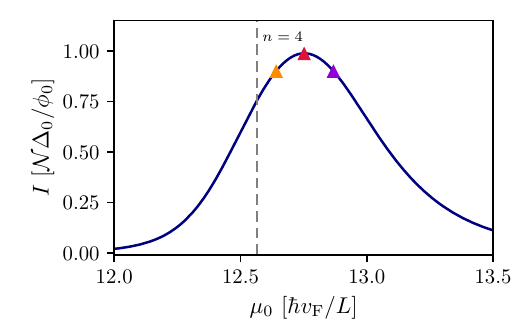}
			\label{fig:sliced_supcurr_vs_kf} 
		\end{minipage}
	}
	\subfloat[]{
		\begin{minipage}[t]{0.47\textwidth}
			\includegraphics[width=\textwidth]{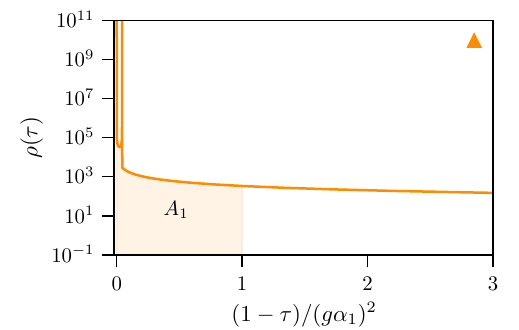}
			\label{fig:rho_vs_tau_kf_1}
		\end{minipage}	
	}\hfill
	\subfloat[]{
		\begin{minipage}[t]{0.47\textwidth}
			\includegraphics[width=\textwidth]{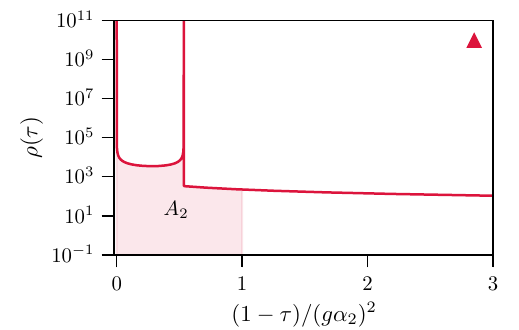}
			\label{fig:rho_vs_tau_kf_2}
		\end{minipage}	
	}
	\subfloat[]{
		\begin{minipage}[t]{0.47\textwidth}
			\includegraphics[width=\textwidth]{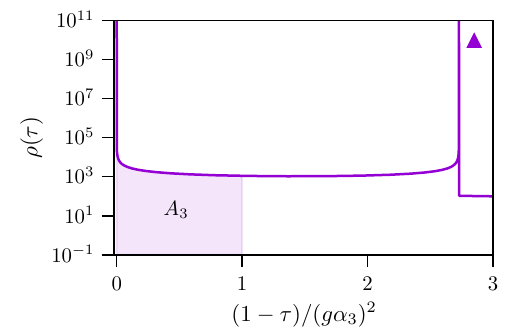}
			\label{fig:rho_vs_tau_kf_3}
		\end{minipage}	
	}
	\caption[]{%
        (a) Magnification around the peak near $n=4$ of Fig.~\ref{fig:I_pi_vs_mu_T_comparison} reported in the main text. The three colored triangles indicate three different value of $\mu_0$ in units of $\hvf/L$, respectively ($12.64,\,12.75,\,12.87$). Other parameters are $\hbar\omega_{\rm r} = 	0.6\,\deltaO$, $T = 0$ and $g=0.1$.
		(b)$-$(d) DOS resolved in the transmission probability calculated at the values of $\mu_0$ indicated by the three colored triangles in \ref{fig:sliced_supcurr_vs_kf}. The supercurrent reaches its maximum at the Fermi level value for which the number of ABSs, whose unperturbed energies fall within an energy interval of order $2\deltaO g\alpha$ (i.e., proportional to the colored regions under the curves), is largest (i.e., the red case). Here, the area ratios are $A_2/A_1\approx1.26$ and $A_2/A_3\approx1.87$.}
	\label{fig:DOS_supcurr_results}
\end{figure*}

In conclusion, to clarify the small shift between the supercurrent peak as a function of $\mu_0$ and the onset of additional values of $k_i$ that fulfill the stationary-wave condition, as shown in Fig.~\ref{fig:I_pi_vs_mu_T_comparison} of the main text, it is helpful to examine the behavior of $\rho(\tau)$.
As a representative case, in Fig.~\ref{fig:DOS_supcurr_results} we show one of the observed peaks in the supercurrent together with $\rho(\tau)$, computed according to Eq.~\eqref{eq:DOS_tau}, for three distinct values of the Fermi level $\mu_{0,l}$ with $l=1,2,3$, each lying slightly above $4 \pi \hbar v_{\rm F}/L$.
These three values correspond, respectively, to the three colored triangles shown in Figs.~\ref{fig:rho_vs_tau_kf_1}$-$\ref{fig:rho_vs_tau_kf_3}.
For each Fermi level, there is an associated value of the photonic mean field, denoted by $\alpha_l$. At that level, we quantify the ABSs whose bare energies satisfy the inequality $ \epsilon(k,\pi)|_{\mu_0=\mu_{0,l}} \leq \deltaO g\alpha_l$, i.e. $\sqrt{1-\tau(k)|_{\mu_0=\mu_{0,l}}}\leq g \alpha_l$. Within the mean-field framework, these correspond to the states that participate in the emergence of the spontaneous TRB instability.
The colored areas in Fig.~\ref{fig:rho_vs_tau_kf_1}$-$\ref{fig:rho_vs_tau_kf_3} emphasized this quantification, 
which we compute as
\begin{equation}
    A_l = \int_{\bar{\tau}_l}^{1} d\tau \rho(\tau)~,
\end{equation}
where $\bar{\tau}_l= 1-(g\alpha_l)^2$. 
This analysis yields $A_2/A_1 \approx 1.26$ and $A_2/A_3 \approx 1.87$, indicating that the optimal condition for achieving a local maximum of the supercurrent exhibits a weak (yet nontrivial) dependence on both the detailed structure of the DOS, close to the value $\tau=1$, and the magnitude of the coupling constant $g$.
In particular, we observe that if the transmission probability, which is close to $\tau = 1$, exhibits a well-defined local minimum inside the narrow range $[1-(g\alpha)^2,1]$, then the DOS acquires a divergence within this same interval, which is clearly separate from the divergence at $\tau = 1$.
As a result of this refined optimization, which goes beyond merely exploiting the divergent behavior of the DOS at $\tau = 1$, the number of ABSs that contribute to instability can be maximized, thus enhancing the supercurrent.

\section{LINEAR RESPONSE FORMALISM AT FINITE TEMPERATURE}\label{app:Linear_response}

In Sec.~\ref{sec:polaritons} of the main-text, we derive the hybridized spectrum of the global system within a linear response formalism.
To this aim, we study Gaussian fluctuations around the mean-field state, by including linear fluctuations on top of the mean-field Hamiltonian, as $\hat{H}_W = \hat{H}_{\rm MF}+\hat{W}$. 
The additional contribution $W$, defined in Eq.~\eqref{eq:W_linear_MF_fluctuations} of the main text, depends on  
the operator $\hat{V}_{\rm F}$, which is introduced in Eq.~\eqref{eq:W_linear_MF_fluctuations_operator} of the main text and is given as a linear combination  
of the operators $\df\hat{H}_{\rm A}$ and $\dqf\hat{H}_{\rm A}$, which were defined in Eq.~\eqref{eq:high_order_term_andreev_continuum} of the main text. Here, rewritten compactly as
\begin{subequations}
	\label{eq:high_order_term_andreev_discrete}
	\begin{align}
		\df\hat{H}_{\rm A} ={\cal N} \int_{-\infty}^\infty \frac{dk}{2\pi} \bigg[P^{z}_k\sigmaz_{k} + P^{x}_k\sigmax_{k}\bigg]~, \label{eq:Paramagn_term_andreev_discrete}\\
		\dqf\hat{H}_{\rm A} ={\cal N} \int_{-\infty}^\infty \frac{dk}{2\pi} \bigg[D^{z}_k\sigmaz_{k} + D^{x}_k\sigmax_{k}\bigg]~.\label{eq:Diamagn_term_andreev_discrete}
	\end{align}
\end{subequations}

To obtain the explicit expression of the term $\hat{V}_{\rm F}$, 
first we insert Eqs.~\eqref{eq:high_order_term_andreev_discrete} in Eq.~\eqref{eq:W_linear_MF_fluctuations_operator} of the main text, then by exploiting the unitary transformation given in Eq.~\eqref{eq:Fermionic_mean_field_quasip}, we obtain 
\begin{equation}
	\hat{V}_{\rm F} = 
    {\cal N} \sum_{j,j'}
    \int_{-\infty}^\infty \frac{dk}{2\pi}
    v_{j,j^{\prime}}(k) 
    \PSidag{j,k}\PSi{j^{\prime},k}~,
	\label{eq:V_linear_MF_fluctuations_operator_full}
\end{equation}
with
\begin{equation}
	\begin{aligned}
	    v_{j,j^{\prime}}(k)  &=\delta_{j,j^{\prime}}\left\{j\left[\Mcos{\theta_k}v_{k}^{z}-\Msin{\theta_k}v_{k}^{x}\right]- v_{0}\right\}+\\
	&+(1-\delta_{j,j^{\prime}})\left[\Mcos{\theta_k}v_{k}^{z}+ \Msin{\theta_k}v_{k}^{x}\right]~,
	\end{aligned}
\end{equation}
where we have defined
\begin{subequations}
				\label{eq:V_linear_MF_fluctuations_operator_full_coeff}
	\begin{align}
		v_{k}^{z} &= gP^{z}_{k} - 2\alpha g^2 D^{z}_{k}~, \label{eq:V_operator_z_coeff}\\
		v_{k}^{x} &= gP^{x}_{k} - 2\alpha g^2 D_{k}^{x}~,\label{eq:V_operator_x_coeff}\\
		v_{0} &= \mathcal{P} - 4\alpha\mathcal{D}~.\label{eq:V_operator_id_coeff}
	\end{align}
\end{subequations}
By definition, the operator $\hat{V}_{\rm F}$ encodes fermionic fluctuations, its mean-field thermal average vanishes, as can be readily verified from Eq.~\eqref{eq:W_linear_MF_fluctuations_operator} of the main text. This implies that the following identity holds
\begin{equation}
	\expval*{\hat{V}_{\rm F}}_{\rm MF} = 
    {\cal N} \sum_j \int_{-\infty}^\infty \frac{dk}{2\pi} 
    v_{j, j}(k)n_{\rm F}(jE(k,\varphi)(\vphi))=0~,
		\label{eq:VF_mean_field_termal_avrg} 
	\end{equation} 
which can be obtained by combining Eq.~\eqref{eq:Fermionic_mean_field_termal_avrg} and Eq.~\eqref{eq:V_linear_MF_fluctuations_operator_full}.

Within the linear response formalism at thermal equilibrium, we consider that the global system is subjected to a general time-dependent, spatially uniform, weak external perturbation 
\begin{equation}
	\hat{V}(t) =f(t)(\B +\Bc) +  {\cal N} \sum_{j,j'} \int_{-\infty}^\infty \frac{dk}{2\pi} 
    f_{j,j^{\prime}}(k,t)
    \PSidag{j, k} \PSi{j^{\prime}, k}~,
	\label{eq:MF_weak_perturbation_general_form}
\end{equation}
where $f(t)$ and $f_{j,j^{\prime}}(k,t)$ denote generic temporal profiles.
Therefore, in the Heisenberg picture, the time evolution of the expectation value of a generic system operator $\hat{O}(t)$ is 
\begin{equation}
	i\hbar\partial_t\expval*{\hat{O}(t)} + \expval*{\comm*{\hat{H}_W}{\hat{O}(t)}}= -\expval*{\comm*{\hat{V}(t)}{\hat{O}(t)}}~.
	\label{eq:MF_linear_response_eq_motion_average}
\end{equation}
Moreover, since we are interested in fluctuations around the mean-field solution, the following decomposition is useful
\begin{equation}
	\expval*{\hat{O}(t)} = \expval*{\hat{O}}_{\rm MF} + \delta\expval*{\hat{O}}~.
	\label{eq:MF_average_fluctuation_scomposition}
\end{equation}
Starting from the expression above, and  applying Eq.~\eqref{eq:MF_linear_response_eq_motion_average}, in combination with Eqs.~\eqref{eq:Bosonic_mean_field_termal_avrg},~\eqref{eq:Fermionic_mean_field_termal_avrg} and~\eqref{eq:VF_mean_field_termal_avrg}, we obtain the following general system of coupled equations
\begin{widetext}
	\begin{equation}
		\begin{cases}
			\begin{aligned}
				&i\hbar\partial_t\big(\delta\expval*{\B}\big)- \big[ \hbar\lambda\omega_{\rm r}\delta\expval*{\B} + \sum_{j,j^\prime}\sum_{\zeta,k}w_{j,j^\prime}(k)\delta\expval*{\PSidag{j,\zeta,k}\PSi{j^\prime,\zeta,k}}\big]= f(t)~,\\
				&i\hbar\partial_t\big(\delta\expval*{\Bc}\big)+ \big[\hbar\lambda\omega_{\rm r}\delta\expval*{\Bc} + \sum_{j,j^\prime}\sum_{\zeta,k} w_{j,j^\prime}(k)\delta\expval*{\PSidag{j,\zeta,k}\PSi{j^\prime,\zeta,k}}\big] = -f(t)~,\\
				&i\hbar\partial_t\big(\delta\expval*{\PSidag{j,\zeta,k}\PSi{j^\prime,\zeta,k}}\big)+(j-j^\prime)E(k,\varphi) \delta\expval*{\PSidag{j,\zeta,k}\PSi{j^{\prime},\zeta,k}} +\big(\delta\expval*{\B}+\delta\expval*{\Bc}\big)w_{j^\prime,j}(k)\big[n_{\rm F}(j^\prime E(k,\varphi)) - n_{\rm F}(j E(k,\varphi))\big]=\\
				&=-f_{j^{\prime},j}(k,t) \Big[n_{\rm F}(j^{\prime}E(k,\varphi)) - n_{\rm F}(jE(k,\varphi))\Big]-\sum_{j^{\prime\prime}}\Big[f_{j^{\prime\prime},j}(k,t)\delta\expval*{\PSidag{j^{\prime\prime},\zeta,k}\PSi{j^{\prime},\zeta,k}}-f_{j^{\prime},j^{\prime\prime}}(k,t)\delta\expval*{\PSidag{j,\zeta,k}\PSi{j^{\prime\prime},\zeta,k}}\Big]~,
			\end{aligned}
		\end{cases}
		\label{eq:MF_linear_response_eq_motion_average_equation_general_system}
	\end{equation}
\end{widetext}
where the matrix elements $w_{j,j^{\prime}}(k)$ are given by
\begin{equation}
	w_{j,j^{\prime}}(k)= \frac{\lambda^{-1/2}}{\sqrt{\mathcal{N}}}v_{j,j^{\prime}}(k)~.
	\label{eq:W_linear_MF_fluctuations_coeff}
\end{equation}
To obtain the system in Eq.~\eqref{eq:MF_linear_response_eq_motion_average_equation_general_system}, we neglect all the terms which are higher than linear in the fluctuations, i.e., terms proportional to $\delta\expval*{\cdot}\delta\expval*{\cdot}$.
For the sake of clarity, from here up to the end of this Appendix, the integration symbol is substituted by a summation over $k$, according to the prescription ${\cal N} \int_{-\infty}^\infty \frac{dk}{2\pi} \to \sum_k$,
and we restore the potential valley degree of freedom $\zeta$.

We next focus on the specific case introduced in Eq.~\eqref{eq:External_weak_flux_probe} of the main text, which can be recast in the equivalent form
\begin{equation}
\hat{V}(t) = \lambda^{-1/2}f(t)(\B+\Bc)+f_\alpha(t)~,
\label{eq:External_weak_flux_probe_2}
\end{equation}
where $
f(t)=\int_{-\infty}^{\infty} \frac{d \Omega}{2 \pi} e^{-i \Omega t} \tilde{f}(\Omega)$ is taken to have no DC component,  $\tilde{f}(0)=0$. The term $f_\alpha(t)=-2{\cal N}\alpha f(t)$ does not contribute to the response and will therefore be disregarded.
Therefore, the linear system given in Eq.~\eqref{eq:MF_linear_response_eq_motion_average_equation_general_system} reduces to
\begin{widetext}
	\begin{equation}
		\begin{cases}
			\begin{aligned}
				&i\hbar\partial_t\big(\delta\expval*{\B}\big)- \big[\hbar\lambda\omega_{\rm r}\delta\expval*{\B} + \sum_{j}\sum_{\zeta,k}w_{j,-j }(k)\delta\expval*{\PSidag{j,\zeta,k}\PSi{-j,\zeta,k}}\big]= \lambda^{-1/2}f(t)~,\\
				&i\hbar\partial_t\big(\delta\expval*{\Bc}\big)+ \big[\hbar\lambda\omega_{\rm r}\delta\expval*{\Bc} + \sum_{j}\sum_{\zeta,k}w_{j,-j }(k)\delta\expval*{\PSidag{j,\zeta,k}\PSi{-j,\zeta,k}}\big] = -\lambda^{-1/2}f(t)~,\\
				&i\hbar\partial_t\big(\delta\expval*{\PSidag{j,\zeta,k}\PSi{-j,\zeta,k}}\big)+j\left[2E(k,\varphi) \delta\expval*{\PSidag{j,\zeta,k}\PSi{-j,\zeta,k}} +\big(\delta\expval*{\B}+\delta\expval*{\Bc}\big)w_{-j,j }(k)\Mtanh{\frac{E(k,\varphi)}{2\kBT}}\right]=0~.
			\end{aligned}
		\end{cases}
		\label{eq:MF_linear_response_eq_motion_average_equation_linearized_system}
	\end{equation}
\end{widetext}
We can compactly express it using a vector notation as follows
\begin{equation}
	\left[i\hbar\partial_t\identity - {\mathcal M}\right]\bm{\delta}\vb{O}(t) = \vb{F}(t)~,
	\label{eq:MF_linear_response_eq_motion_average_equation_linearized_system_2_matrix}
\end{equation}
where we define two vectors
\begin{equation}
	\begin{aligned}
		\bm{\delta}\vb{O}(t) &= \left[\delta\expval*{\B}, \delta\expval*{\Bc},\delta\expval*{\PSidag{j,\zeta,k}\PSi{-j,\zeta,k}},\dots\right]^{T}~,\\
		\vb{F}(t) &= \left[\lambda^{-1/2}f(t), -\lambda^{-1/2}f(t),0, \dots\right]^{T}~.
	\end{aligned}
\end{equation}
Then, we define the Green's function relative to the system of linear differential equations in Eq.~\eqref{eq:MF_linear_response_eq_motion_average_equation_linearized_system_2_matrix} as
\begin{equation}
	\left[i\hbar\partial_t\identity - {\mathcal M}\right]{\mathcal G}(t) = \delta(t)~.
	\label{eq:MF_linear_response_eq_motion_green_funct}
\end{equation}
%
In particular, our focus is on the retarded Green's function, which enables the expression of the response of the system to the perturbation as
\begin{equation}\label{eq:LRT-t}
\bm{\delta}\vb{O}(t) = \int^t_{-\infty} d t' {\cal G}^{\rm R}(t-t') \vb{F}(t')~.
\end{equation}
Upon applying the Fourier transform
\begin{equation}
{\mathcal G}^{\rm R}(t)=\int_{-\infty}^\infty \frac{d\Omega}{2\pi} e^{-i \Omega t}
\tilde{\mathcal G}^{\rm R}(\Omega)~,
\end{equation}
it is convenient to formulate the problem in the frequency domain, 
which leads to  
\begin{equation}
 \tilde{\mathcal G}^{\rm R}(\Omega)=\left[(\hbar \Omega+i 0^+)\identity - {\mathcal M}\right]^{-1}~, 
\label{eq:MF_linear_response_eq_motion_retarded_green_funct}
\end{equation}
thus, we represent Eq.~\eqref{eq:LRT-t} within the frequency domain as
\begin{equation}
	\bm{\delta}\tilde{\vb{O}}(\Omega) = \tilde{{\mathcal G}}^{\rm R}(\Omega)\tilde{\vb{F}}(\Omega)~.
	\label{eq:MF_linear_response_eq_motion_green_funct_freq}
\end{equation}
It is useful to represent $\tilde{{\mathcal G}}^{\rm R}(\Omega)$ in a block form as
\begin{equation}
	\tilde{\mathcal G}^{\rm R}(\Omega)
	\equiv \mqty[ 	
    \tilde{\mathcal G}^{\rm R}_{\rm BB}(\Omega) & \tilde{\mathcal G}^{\rm R}_{\rm FB}(\Omega)\\  	
    \tilde{\mathcal G}^{\rm R}_{\rm BF}(\Omega)& \tilde{\mathcal G}^{\rm R}_{\rm FF}(\Omega)
    ]~, 
	\label{eq:MF_linear_response_eq_motion_green_funct_freq_blocks}
\end{equation}
where the labels B and F refer, respectively, to fluctuations in the photon (bosonic) sector and to fluctuations in the Andreev subspace (fermionic sector).

To evaluate the response of the coordinate of the LC quantum oscillator, it suffices to evaluate only the $2\times2$ matrix $\tilde{\mathcal G}^{\rm R}_{\rm BB}$. 
To this end, we express the matrix ${\cal M}$ in block form and by exploiting the block matrix inversion formula, we obtain%
\begin{equation}
\begin{aligned}
\tilde{\mathcal G}^{\rm R}_{\rm BB}(\Omega) &=
\{(\hbar \Omega+i 0^+)\identity_{\rm BB}-{\cal M}_{\rm BB}\\
&-{\cal M}_{\rm BF}[(\hbar \Omega+i 0^+)\identity_{\rm FF}-{\cal M}_{\rm FF}]^{-1}{\cal M}_{\rm FB}\}^{-1}~,
\end{aligned}
\end{equation}
which is given explicitly by
\begin{widetext}
\begin{equation}
	\begin{aligned}
		\tilde{\mathcal G}^{\rm R}_{\rm BB} (\Omega)= \mqty[\hbar\Omega+i0^+ - \hbar \lambda\omega_{\rm r} -\lambda^ {-1}\tilde{\chi}(\Omega)  & -\lambda^ {-1}\tilde{\chi}(\Omega) \\ \lambda^ {-1}\tilde{\chi}(\Omega) & \hbar\Omega+i0^+ + \hbar \lambda\omega_{\rm r}+\lambda^ {-1}\tilde{\chi}(\Omega)]^{-1}~,
	\end{aligned}
\end{equation}
\end{widetext}
where $\tilde{\chi}(\Omega)$ is reported in Eq.~\eqref{eq:MF_linear_response_chi} of the main-text.
%
Employing the decomposition in Eq.~\eqref{eq:MF_average_fluctuation_scomposition}, the response of the coordinate of the LC quantum oscillator, $\hat{X}=\A + \Ac$, reads as
\begin{equation}
\delta X(t)  =\lambda^{-1/2}(\delta\expval*{\B}+ \delta\expval*{\Bc})~.
\end{equation}
Therefore, in frequency domain, from Eq.~\eqref{eq:MF_linear_response_eq_motion_green_funct_freq} we obtain the result reported in Eq.~\eqref{eq:MF_linear_response_eq_motion_coord_freq} of the main text
\begin{equation*}
\delta\tilde{X}(\Omega)=\tilde{\Pi}(\Omega)\tilde{f}(\Omega)~,
\end{equation*}
where $\tilde{\Pi}(\Omega)$ is expressed in terms of all elements of the $2\times 2$ matrix $[\tilde{\mathcal G}^{\rm R}_{\rm BB}(\Omega)]_{\ell,\ell'}$ ($\ell,\ell' \in \{1,2\}$) as
\begin{equation}
	\begin{aligned}
	\tilde{\Pi}(\Omega)&= \lambda^{-1}\big\{[	\tilde{\mathcal G}^{\rm R}_{\rm BB}(\Omega)]_{11} - [	\tilde{\mathcal G}^{\rm R}_{\rm BB}(\Omega)]_{12} \\
    &+ [	\tilde{\mathcal G}^{\rm R}_{\rm BB}(\Omega)]_{21} - [	\tilde{\mathcal G}^{\rm R}_{\rm BB}(\Omega)]_{22}\big\}~.
	\end{aligned}
\end{equation}
In conclusion, we emphasize that the zeros of the determinant 
\begin{equation}
	\rm{det}([\tilde{\mathcal G}^{\rm R}_{\rm BB}(\Omega)]^{-1})= (\hbar\Omega+i0^+)^2  - (\hbar\lambda\omega_{\rm r})^2 - 2\hbar\omega_{\rm r}\tilde{\chi}(\Omega)~,
\end{equation}
coincide with the poles of the response function $\tilde{\Pi}(\Omega)$, explicitly reported in Eq.~\eqref{eq:MF_linear_response_ret_response_funct_freq} of the main-text, and those represent the low-energy spectrum of collective hybridized light-matter excitations.
%

\bibliography{biblio}

\end{document}